\documentclass[aps,floatfix,notitlepage, tightenlines, nofootinbib]{revtex4-1}
\usepackage{amsmath, amssymb, amsfonts, amsthm, latexsym, epsfig, mathrsfs, xcolor, bbm, slashed}
\usepackage[top=1in, bottom=1in, left=.9in, right=.9in]{geometry}
\numberwithin{equation}{section} % JHEP style eqn numbering

\usepackage{mdframed}
\DeclareMathAlphabet{\mathpzc}{OT1}{pzc}{m}{it}
\usepackage[cal=boondox, calscaled=.94, bb=ams, scr=rsfs]{mathalfa} 
\usepackage[T1]{fontenc} % for thorn and eth
\usepackage{amsfonts}
\usepackage{graphicx}
\usepackage{amsmath}
\usepackage{color}

\usepackage[colorlinks=true, citecolor=blue,urlcolor=blue]{hyperref}
\usepackage{hyperref}
\usepackage{PeterStyle} 
\usepackage{subfigure}
\usepackage{comment}
\usepackage{multirow}
\usepackage{calligra}

\newcommand{\he}{\hat{e}}

\newcommand{\hp}{{h_+}}
\newcommand{\hm}{{h_-}}

\newcommand{\Dp}{{\Delta_+}}
\newcommand{\Dm}{{\Delta_-}}

\newcommand{\adst}{\mathrm{AdS}_2}
\newcommand{\adsf}{\mathrm{AdS}_5}
\newcommand{\ads}[1]{\mathrm{AdS}_{#1}}
\newcommand{\rnads}{pRN-AdS${}_5$}
\newcommand{\Rin}{{R}^{\rm in}}

\newcommand{\Ru}{{\underline R}}
\newcommand{\G}[1]{\Gamma(#1)}
\def\pd{\partial}
 
\newcommand{\xih}{\hat{\xi}}
\newcommand{\Au}{{\underline A}}
\newcommand{\Su}{{\mathcal{\underline S}}}

\newcommand{\OO}{\mathcal{O}}
\begin{document}

\title{Semi-local Quantum Criticality and the Instability of Extremal Planar Horizons}
\author{Samuel E.~Gralla\footnote{{\tt sgralla@email.arizona.edu}}}
\author{Arun~Ravishankar\footnote{{\tt arunravishankar@email.arizona.edu}}}
\author{Peter~Zimmerman\footnote{{\tt peterzimmerman@email.arizona.edu}}}
\affiliation{Department of Physics, University of Arizona}
\date{\today}
\begin{abstract}
We show that the Aretakis instability of compact extremal horizons persists in the planar case of interest to holography and discuss its connection with the emergence of ``semi-local quantum criticality'' in the field theory dual.
In particular, the spatially localized power-law decay of this critical phase corresponds to spatially localized power-law \textit{growth} of stress-energy on the horizon.  For near-extremal black holes these phenomena occur transiently over times of order the inverse temperature.  The boundary critical phase is characterized by an emergent temporal conformal symmetry, and the bulk instability seems to be essential to preserving the symmetry in the presence of interactions.  We work primarily in the solvable example of charged scalar perturbations of five-dimensional (near-)extremal planar Reissner-Nordstr\"om anti-de Sitter spacetime and argue that the conclusions hold more generally.
\end{abstract}
\maketitle
\newpage
\tableofcontents
\newpage
\section{Introduction}\label{sec:intro}

With the advent of holographic duality, the problem of black hole stability (and instability) is attaining an increasingly wide importance in physics.  Aside from its continuing relevance to astrophysics and quantum gravity, black hole perturbation theory has gained new life as a means of defining and exploring certain \textit{non-gravitational} field theories.  Notably, the ``AdS/CMT correspondence''  \cite{Hartnoll:2016apf} aims to engineer models that describe laboratory  phenomena such as strange metallicity \cite{strangemetals,Faulkner:2010zz,Iqbal:2011ae}.  In the new science of black holes, establishing gravitational stability can help qualify a field theory for laboratory relevance, while finding an instability can point to a new discovery in fields as far-flung as astrophysics and condensed matter.

A very interesting development occurred in 2010 when the mathematician Aretakis discovered a rather subtle instability of the extremal Reissner Nordstr\"om (RN) spacetime \cite{Aretakis:2010gd,Aretakis:2011ha,Aretakis:2011hc}.  He considered massless scalar perturbations and managed to prove that the field and all its derivatives decay everywhere outside the event horizon.  However, he showed that something unusual happens precisely \textit{on} the horizon: while the field decays, the first radial  derivative does not.  Furthermore, the second derivative grows linearly in time, and higher derivatives grow at higher polynomial rates: 
\begin{align}\label{Aretakis}
    (\pd_r)^n\Phi|_{\mathcal{H}} > C v^{n-1}.
\end{align}
Here $\Phi$ is the scalar field, $v$ is ingoing time (equal to affine time on the future degenerate horizon $\mathcal{H}$), $r$ is the areal radius, $C$ is some constant, and $n \geq 2$ derivatives are taken.  The original result \eqref{Aretakis} held only for initial data that was non-zero on the event horizon, but it soon became clear that the instability persists for data supported arbitrarily far away \cite{Lucietti:2012xr,Aretakis:2012bm,Casals:2016mel}, although sometimes with different rates \cite{Angelopoulos:2018uwb}.  The result was also extended to near-extremal black holes, where the growth occurs transiently (for a time of order the inverse temperature) near the horizon \cite{Lucietti:2012xr,Murata:2013daa,Gralla:2016sxp,Zimmerman:2016qtn}.  
Although the precise power laws differ from example to example, this typical behavior of successive derivatives growing one power of $v$ faster has now been seen in a wide variety of perturbation problems involving (near-)extremal black holes, and is known as the Aretakis instability.

This development is especially intriguing for holography, as many of the interesting phenomena driving the AdS/CMT correspondence occur in the low-temperature (extremal) limit. 
However, these holographic models generally contain \textit{planar} black holes (so that the dual theory can live in Minkowski space), while the Aretakis instability has thus far only been exhibited for compact horizons.  Intuitively, it is not clear that the instability will persist in the non-compact case, as one might imagine that the perturbation will spread out and decay.  Furthermore, general proofs of the instability \cite{Aretakis:2011ha,Aretakis:2011hc,aretakis2012decay,Aretakis:2012ei,Lucietti:2012sf,Aretakis:2012bm} rely in an essential way on the assumption of compact horizon cross-sections.  However, our recent work identifying the instability with $\adst$-factored near-horizon geometries \cite{gralla2018scaling} suggests that it will remain, since extremal planar black holes do possess such limits.

In this paper we will show that the instability does in fact persist for certain planar horizons and argue that it persists generically.  In fact, it is closely related to a phenomenon already understood in the condensed matter \cite{Varma:1989zz,2000Natur.407..351S,2001Natur.413..804S} and AdS/CMT literature \cite{Iqbal:2011in,Iqbal:2011ae}, termed \emph{semi-local quantum criticality}. 
We mainly consider the well-studied model of a charged scalar field perturbing a  planar Reissner-Nordstr\"om black hole in five-dimensional Anti-de Sitter space (\rnads), which defines a finite-density field theory on 3+1 dimensional Minkowski space.  We denote the three Cartesian directions by $\vec{y}$ and use an ingoing time coordinate $V$ that labels ingoing null geodesics by their time $t$ in the boundary theory (in particular, $V\to t$ at the boundary $r \to \infty$).  The model is characterized by two scales, which can be taken to be the temperature $T$ and the chemical potential $\mu$.  In the low-temperature limit $T \ll \mu$, we may consider the regime
\begin{align}\label{regime}
\quad 1/\mu \ll V-V' \ll 1/T, \qquad 1/\mu \ll |\vec{y}-\vec{y}'| \ll V-V',
\end{align}
where we set $\hbar=k_b=c=1$.  The first assumption is sufficient to see the critical behavior and Aretakis growth,  while the second allows for a particularly transparent mathematical expression that we focus on in this introduction.

We impose Dirichlet boundary conditions and consider initial data supported near the boundary $r \to \infty$. The field response is encoded by the boundary-to-bulk propagator $G_{\rm \pd B}$, defined in Eq.~\eqref{eq:G Bpd def} below.  Restricting to a regime of parameter space where there are no exponential instabilities, we show that this propagator may be approximated in the regime \eqref{regime} by 
\begin{align}\label{money}
    G_{\pd \rm B} 
    \approx \mathcal{A} \ell^{-3/2}\, \frac{e^{-y/\xi}}{(\mu y)^4} \left(\frac{y}{\xi}\right)^{3/2} \begin{cases} Z(r) (\mu V)^{-1}, & r \gg 1/V \ \textrm{(far-zone)} \\ \left[ \, \mu V(1+\sqrt{6}(r/r_0-1)V \mu) \right]^{-1/2}, & |r-r_0| \ll r_0 \ \textrm{(near-zone)} \end{cases},
\end{align}
where the function $Z(r)$ and numerical constant $\mathcal A$ may be inferred from Eqs.~\eqref{eq:G pd large x} and \eqref{eq:G pdpd tail} below.\footnote{Alternatively one may refer to Eqs.~\eqref{eq:G near pd qnm} and \eqref{eq:G pdpd tail T} at early times.}  The ``correlation length'' $\xi \sim 1/\mu$ is a length scale set by near-horizon physics; the precise formula is $\xi=\sqrt{6}\hat{\xi}/\mu$ using Eq.~\eqref{eq:xihat} below.

\begin{figure}
 \centering
    \includegraphics[width=\textwidth]{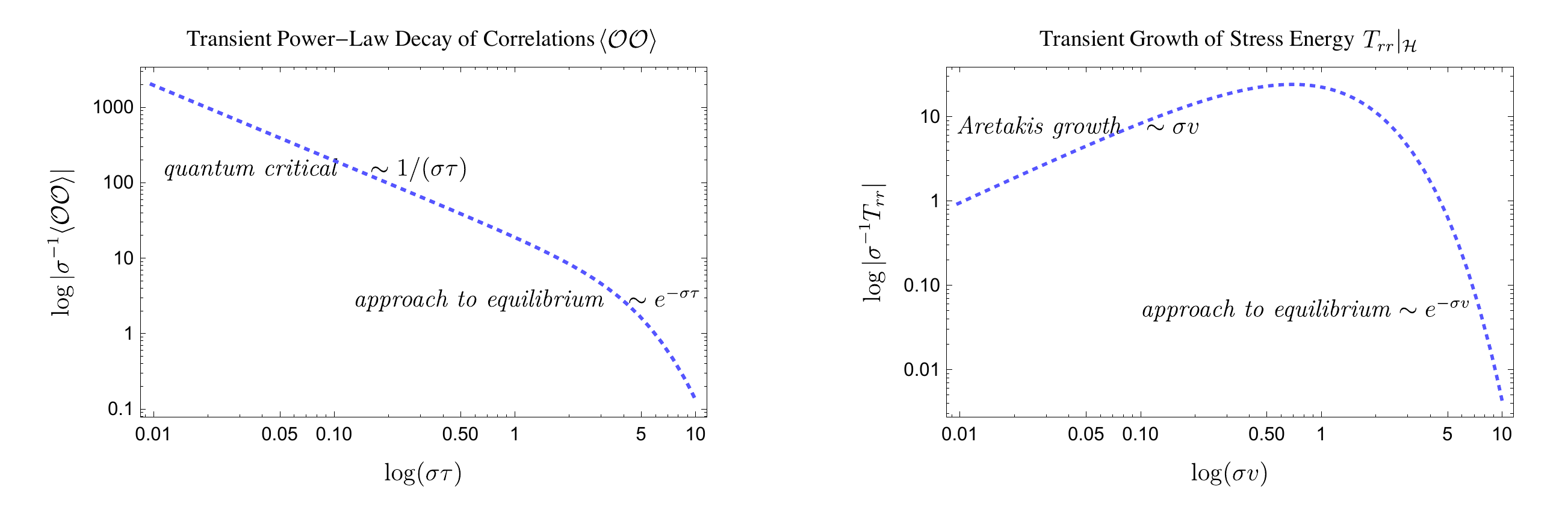}
    \caption{The quantum critical behavior of the dual theory (left) is associated with growth of bulk stress-energy on the horizon (right).  We use a dimensionless temperature %$\sigma=\pi\ell^2 T/(3r_0)$,
    $\sigma=\pi \mu T/(3\sqrt{6})$,
    boundary time $\tau=\mu t/\sqrt{6}$, and bulk advanced time $v=\mu V/\sqrt{6}$.  We choose parameters in the range \eqref{nopoles}.   The left panel shows the boundary retarded two-point function, which decays like $1/\tau$ in the quantum critical regime \eqref{regime} before exponentially approaching equilibrium when $\tau \gtrsim \sigma$.  The right panel shows the scalar stress energy component $T_{rr}\sim \abs{\pd_r\Phi}^2$ on the horizon, which grows linearly in $v$ before similarly decaying.  In the extremal limit, the power-law behaviors persist forever.}
    \label{fig:regimes}
\end{figure} 

Equation~\eqref{money} provides a late-time approximation valid all the way from the boundary $r=\infty$ to the horizon $r=r_0$.  Examining these limiting values will reveal the semi-local quantum critical behavior  and the Aretakis instability, respectively.  At the boundary, the dual two-point function is defined by peeling off the leading power of $r$ [Eq.~\eqref{eq:G pdpd def} below], giving
\begin{align}\label{eq:OO tail}
 G_{\pd\pd} = \langle \mathcal O(t,\vec{y}) \mathcal O(0,0) \rangle \approx \mathcal B\, 
 \mu^{-5} \frac{e^{-y/\xi}}{y^4}\left(\frac{y}{\xi}\right)^{3/2}\frac{1}{t},
\end{align}
where $\mathcal B$ is a dimensionless constant (equal to $36\sqrt{6}$ times the quantity $\mathcal C_f$ given in Eq.~\eqref{eq:C pdpd} below) 
and where by $\approx$ we refer to the regime \eqref{regime}.  This is the semi-local critical behavior: there is a finite correlation length $\xi$ in space, but an infinite correlation length in time ($1/t$ decay).  

On the other hand, from the near-zone expression in \eqref{money} we see the Aretakis instability on the horizon,
\begin{align}\label{eq:pdnGBpd}
    (r_0\pd_r)^n G_{\pd \rm B}|_{r=r_0} \approx \ell^{-3/2}\mathcal{A}_n\, \frac{e^{-y/\xi}}{(\mu y)^4}\left(\frac{y}{\xi}\right)^{3/2} (\mu V)^{n-1/2},
\end{align}
 where $\mathcal{A}_n$ is an overall numerical factor depending on the number of derivatives.  We see that each derivative has its own version of semi-local criticality, with derivative-dependent exponent such that $n\geq 1$ entails growth.  In particular, the stress-energy of the field will grow linearly in time. 

 The critical behavior and Aretakis growth can both be understood in terms of scaling symmetry.  In Eq.~\eqref{money}, the far-region scales as $\lambda$ under $V \to \lambda^{-1} V$, while the near-region scales as $\lambda^{1/2}$ under $V \to \lambda^{-1} V$ \textit{and} $r-r_0 \to \lambda(r-r_0)$.  The former scaling accounts for the quantum critical behavior \eqref{eq:OO tail}, while the latter scaling accounts for the Aretakis growth \eqref{eq:pdnGBpd}.  This second scaling is familiar from near-horizon limits, and indeed the Aretakis instability can indeed be generally understood as self-similarity under the flow to late times near the horizon that yields the near-horizon geometry \cite{Gralla:2017lto}.  We discuss these scalings in more detail in Sec.~\ref{sec:interactions} below.
 
 The critical behavior and Aretakis growth last forever for an extremal black hole, but in the near-extremal case they are cut off at a time of order $1/T$, after which perturbations decay exponentially.  We have been able to analytically exhibit this behavior using techniques developed in \cite{Gralla:2016sxp} (Sec.~\ref{sec:less extreme} below).  Fig.~\ref{fig:regimes} illustrates these features.
 
We thus see that the semi-local critical behavior in the dual theory is intimately related to the Aretakis instability in the bulk.  Indeed, we may say that the Aretakis instability \textit{is} semi-local critical behavior on the horizon, with growth (e.g. of stress-energy) rather than decay.  What does this growth mean for the dual theory?  One important point is that, organizing the instability in terms of the scaling symmetry, it follows straightforwardly that all scalar invariants decay \cite{Gralla:2017lto,Hadar:2017ven,Burko:2017eky}.   This makes the Aretakis growth of the ``weak null'' type visible only to infalling observers.  Thus identifying a CFT dual description will be quite challenging, as the dual description of infalling observers is generally mysterious.  In particular, we argue in Sec.~\ref{sec:interactions} that the symmetry also ensures that dual correlation functions all decay, independently of the type of interactions included.  We give some discussion of the potential holographic meaning of the instability, but in general leave this question to future work.

While we have worked with a particular spacetime, we expect the Aretakis instability to be present for generic planar event horizons.  In fact, in a certain sense we expect the instability to be \textit{stronger} than in compact-horizon spacetimes.  A universal formulation \cite{Gralla:2018xzo} of the Aretakis effect is that, quite generally, fields near extremal horizons take the form $V^{-h}f(V(r-r_0))$ for some exponent $h$ satisfying $\textrm{Re}[h] \geq 1/2$. 
For compact horizons, each angular mode of the field has its own exponent, and the late-time behavior is dominated by the one with the smallest real part.  Depending on the details, this could be a quite large value of $h$, requiring several derivatives to see any growth.  However, for non-compact horizons, there is instead a continuum of values, with the minimum $h=1/2$ generically making some contribution.  (This arises in the example we study and also for many other models considered in the AdS/CMT context \cite{Hartnoll:2016apf}.\footnote{The correlation length $\xi$ is the spatial wavelength associated with the minimal value $h=1/2$ and sets the typical distance scale above which the $h=1/2$ behavior is dominant.  However, the $h=1/2$ behavior is present at all spatial distances, and growth of a single derivative always occurs for sufficiently low temperatures at any spatial point.})  We therefore conclude that, like the example in this paper, the Aretakis instability of non-compact extremal horizons will generically require only a single derivative for growth.

The remainder of this paper is organized as follows. Sec.~\ref{sec:RN-AdS} presents parameter, gauge, and coordinate definitions used to describe the background solution in simple holographic terms.  In Sec.~\ref{sec:charged scalar} we decompose a charged, massive scalar perturbation into a sum over Fourier modes to formulate the perturbation equation as a radial ODE for a given mode. 
In Sec.~\ref{sec:extremal} we review the zero-temperature, low-frequency solutions and study the associated late-time behavior of perturbations.  We then consider nearly-extremal black holes in Sec.~\ref{sec:less extreme}, showing how the power-law behavior manifests in a coherent sum of quasinormal modes.  Finally, in Sec.~\ref{sec:interactions} we discuss interactions and argue that dual correlators will not display growth.

\section{Background configuration}\label{sec:RN-AdS}

A Maxwell field coupled to AdS gravity in five dimensions has bulk action
\begin{equation}\label{eq:S}
S=\int\left(R+\frac{12}{\ell^2}-\frac{1}{4}F^2\right)\sqrt{-g} \,d^5x,
\end{equation}
where we now set $G=1$ in addition to $\hbar=k_B=c=1$.  The AdS length $\ell$ is related to the cosmological constant $\Lambda_c$ by  $\ell^2=-6/\Lambda_c$. 
We consider the planar charged black hole RN-$\ads{5}$ solution \cite{Chamblin:1999tk}, which may be written
\begin{align}\label{eq:RN-AdS stuff}
ds^2&=-Ndt^2+N^{-1}dr^2+(r^2/\ell^2)d\vec{y}^2,\qquad 
N=\frac{(r^2-r_0^2)\Big(r^4+r^2r_0^2-2r_0^4(1-3\sigma)\Big)}{r^4 \ell^2},\\
A&=\frac{r_0\sqrt{6(1-3\sigma)}}{\ell^2}(1-(r_0/r)^2) dt.
\end{align}
The boundary is $r\to \infty$ and the event horizon is at $r=r_0$.  The solution contains three parameters ($r_0,\sigma,\ell$), but one can be removed by a diffeomorphism.  The relevant physical scales for the boundary theory are the temperature $T$ and chemical potential $\mu$ \cite{Faulkner:2009wj,Hartnoll:2016apf}, given by 
\begin{align}\label{eq:T and mu}
    T=\frac{ r_0}{\ell^2 } \frac{3}{\pi}\sigma \qquad
    \mu=\frac{r_0}{\ell^2}\sqrt{6(1-3\sigma)}.
\end{align}

The dimensionless boundary theory quantity is the ratio,
\begin{align}
    \frac{T}{\mu}=\frac{3}{\pi}\frac{\sigma}{\sqrt{6(1-3\sigma)}}.
\end{align}
Notice that $T/\mu\propto \sigma$ in the extremal limit $\sigma \to 0$.

We now remove $r_0$ by introducing dimensionless coordinates defined by
\begin{align}\label{eq: scaled coords}
    z=\frac{r_0}{r}, \qquad \tau = \frac{t}{\ell^2/r_0}, \qquad  x = \frac{y}{\ell^2/r_0}.
    %y = \frac{x}{\ell^2/r_0}.
\end{align}
Eq.~\eqref{eq:RN-AdS stuff} becomes
\begin{equation}\label{eq:metric tau}
    ds^2=\frac{\ell^2}{z^2}(-fd\tau^2+dz^2/f+ d\vec x^2),\qquad A=\sqrt{6(1-3\sigma)}(1-z^2)d\tau,
\end{equation}
where \begin{equation}\label{eq:f}
    f=1-3z^4+2z^6.
\end{equation} 
The event horizon is at $z=1$ and the boundary is at $z=0$.  These coordinates and gauge are not regular on the horizon.  A set of regular coordinates and gauge are given by
\begin{align}\label{eq:ingoing defs}
    dv=d\tau-\frac{dz}{f}, \quad A'=A+d\left(-f^{-1}\sqrt{6(1-3\sigma)}(1-z^2)\right).
\end{align}
We fix the integration constant so that $v=\tau$ on the boundary $z=0$.  The solution now becomes
\begin{equation}\label{eq:ingoing met and A}
    ds^2=\frac{\ell^2}{z^2}(-fdv^2-2dvdz+d\vec x^2),\qquad A'=\sqrt{6(1-3\sigma)}(1-z^2)dv.
\end{equation}
These coordinates are convenient for calculations, while the original coordinates are convenient for the holographic dictionary \eqref{eq:T and mu}.  Note that the dimensional ingoing coordinate $V$ used in the introduction
which reduces to the boundary time $t$ is related to $v$ by \[ V = \frac{\ell^2 v}{r_0}. \]   This coordinate labels null geodesics emanating inward from the boundary at time $t$.

\section{Charged scalar perturbation}\label{sec:charged scalar}

As a perturbation, we take a massive charged scalar $\Phi$ satisfying
\begin{equation}\label{eq:u1 scalar eq}
(D^2-m^2)\Phi=0, \qquad D=\nabla-iqA.
\end{equation}
The mass $m$ and charge $q$ are at this stage classical inverse length scales, as we set $G=c=1$ but not $\hbar=1$.
The field $\Phi$, as determined from suitable initial/boundary data or a compact source and boundary data, may be constructed from the retarded two-point function $G$ satisfying 
\begin{equation}\label{eq:G5}
(D^2-m^2)G=\delta_5,
\end{equation}
where $\delta_5$ is the invariant delta distribution $\delta_5(x^\mu,x^\mu{}')=\delta^5(x^\mu-x^\mu{}')/\sqrt{-g}$.  We set $t'=0$ and $x'=0$ using the background symmetries.\footnote{In what follows we will also use ingoing time, but the difference $t'-v'$ is zero at radial infinity where take the source point in the final calculation.} Working in frequency space for the non-radial directions,
\begin{equation}\label{eq:G5 modes}
    G=\int \frac{d\omega d^3k}{(2\pi)^4} e^{-i\omega \tau+i\vec{k}\cdot\vec{x}} g(\omega,k;z,z').
\end{equation}
We refer to $g(z,z')$ as the ``transfer function'' for each mode. The transfer function obeys the inhomogenous radial equation 
\begin{equation}\label{eq:trans eq}
g''+\left(\frac{f'}{f}-\frac{3}{z}\right)g'+\left(\frac{(\omega+\ell q A_\tau)^2}{f^2}-\frac{\vec{k}^2}{f}-\frac{\ell^2 m^2}{z^2f}\right)g= \frac{z^3}{\ell^3 f}\delta(z-z'),
\end{equation}
where prime denotes differentiation with respect to $z$.  Notice that the mass and charge appear only in the dimensionless combinations $\ell q$ and $\ell m$.  We denote the homogeneous solutions by $R$,
\begin{equation}\label{eq:R eq}
R''+\left(\frac{f'}{f}-\frac{3}{z}\right)R'+\left(\frac{(\omega+\ell qA_\tau)^2}{f^2}- \frac{\vec{k}^2}{f}-\frac{\ell^2 m^2}{z^2 f}\right)R =0.
\end{equation}
Near the boundary $z=0$, the general solution is a linear combination of $z^{\Delta_+}$ and $z^{\Delta_-}$, where
\begin{align}\label{eq:Delta}
\Delta_\pm=2\pm\sqrt{4+\ell^2m^2}.
\end{align}
We define solutions $R^\pm$ distinguished by having a single behavior on the boundary (with unit normalization),
\begin{align}\label{eq:Rpm asy}
    R^{\pm} \sim z^{\Delta_\pm}, \qquad z \to 0.
\end{align}
Near the horizon the general solution is a linear combination of ingoing and outgoing waves, with the precise asymptotics differing in the extremal and non-extremal cases (see Secs.~\ref{sec:extremal} and \ref{sec:less extreme} below).  We denote the solution with pure ingoing waves at the horizon by $\Rin$.  

We construct the transfer function from these two homogeneous solutions by
\begin{equation}\label{eq:g def}
    g(z,z')=\frac{\Rin(z_>)R^+(z_<)}{\mathcal W},
\end{equation}
where $z_{<}=\min(z,z')$, $z_>=\max(z,z')$ and $\mathcal W=\ell^3 z^{-3} f(z) \mathbf (R^+\pd_z \Rin-\Rin \pd_z R^+)$ is a constant.  This corresponds to choosing the retarded Green function with ``Dirichlet'' conditions at the boundary.  

\subsection{Boundary behavior}
Above we have treated classical propagation of fields in the bulk spacetime.  With holography in mind, it is convenient to define propagators where one or both points is taken the boundary, with the leading behavior $z^{\Delta_+}$ peeled off.  In particular, we introduce the bulk-boundary propagator as
\begin{align}\label{eq:G Bpd def}
    G_{\pd \rm B}(\tau,x,z) := \ell^{3/2} \lim_{z'\to0}{(z')}^{-\Delta_+} G 
\end{align}
and the boundary-boundary propagator as
\begin{align}\label{eq:G pdpd def}
     G_{\pd \pd}(\tau,x) & := \ell^{3} \lim_{z \to 0} \lim_{z'\to0}{(z z')}^{-\Delta_+} G \\ & = \ell^{3/2}\lim_{z \to 0} {(z )}^{-\Delta_+} G_{\pd \rm B}.
\end{align}

As it stands, these are just useful devices for keeping track of sources and fields near the boundary in classical propagation of fields.  However, $G_{\pd \pd}$ is intrinsically 3+1-dimensional and hence is a natural candidate for a propagator in a dual theory.  Indeed, we find in Sec. \ref{far tail stuff} below that $G_{\pd \pd}$ does agree (at least in the regime we consider) with the retarded two-point function for the dual operator as normally defined \cite{Son:2002sd}.

\section{Extremal case}\label{sec:extremal}

We begin with the precisely extremal case $\sigma=0$.  The radial equation \eqref{eq:R eq} can be solved for $\omega \ll 1$ using the method of matched asymptotic expansions  \cite{Faulkner:2009wj}.  The regions of interest are
\begin{align}\label{eq:MAE}
    \textrm{near-region: } & 1 - z \ll 1 \\
    \textrm{far-region: } & 1 - z \gg \omega \\
    \textrm{overlap region: } & \omega \ll 1-z \ll 1.
\end{align}

\subsection{Near-region}

The near equation is Eq.~\eqref{eq:R eq} with $\omega \to 0$ fixing $\omega/(1-z)$.  On general grounds \cite{Gralla:2018xzo} this must produce the equation for a massive, charged scalar on $\ads{2}$, and indeed we find
\begin{align}\label{near}
    ((1-z)^2R')'+\left(\left(\frac{\omega}{12(1-z)}+\hat e\right)^2 -\hat m^2\right)R = 0, 
\end{align}
where the effective mass $\hat{m}$ and charge $\hat{e}$ are
\begin{equation}\label{eq:ads2 charge and mass}
    \he=\ell q/\sqrt{6},\qquad\hat{m}^2=\frac{1}{12}(k^2+\ell^2 m^2).
\end{equation}
Following \cite{Gralla:2018xzo}, it is convenient to discuss three different solutions,\footnote{The headgear on the $\pm$ is intentional; this distinguishes the solutions $R^{\widehat{\pm}}$ (defined based on overlap region behavior) from $R^{\pm}$ (defined based on $\ads{5}$ boundary behavior).  The notation $R^{\pm}_{\rm near}$ would mean the near limit of the full solution $R^{\pm}$.  Note that there are some subtleties at certain discrete values of $h_{\pm}$, which are described in detail in Sec.~3 of Ref.~\cite{Gralla:2018xzo}.  In the present work, these cases are measure-zero and can be ignored.}
\begin{align}\label{eq:R near  pmhat}
R^{\widehat{\pm}}_{\rm near} = (-i\omega/6)^{-h_\pm}M_{i\he,h_\pm-1/2}\left(-\frac{i\omega}{6(1-z)}\right), \qquad R_{\rm near}^{\rm in} = W_{i\he,h_+-1/2}\left(-\frac{i\omega}{6(1-z)}\right),
\end{align}
where we introduce
\begin{align}\label{eq:h}
h_\pm = 1/2\pm\nu, \qquad \nu=\sqrt{1/4+\hat m^2-\he^2}.
\end{align}
These are the scaling dimensions of $\ads{2}$ holography.  In particular, as $z \to 1$ (the $\ads{2}$ boundary or overlap region) the solutions behave as a linear combination of $(1-z)^{-h_+}$ and $(1-z)^{-h_-}$.  %$\ell_2$ is the $\ads{2}$ length scale $\ell_2^2 = \ell^2/12$.
The solutions $R^{\widehat{\pm}}_{\rm near}$ have just one overlap behavior and are normalized to unity,
\begin{align}\label{eq:R wide hat pm}
    R^{\widehat{\pm}}_{\rm near} \sim (1-z)^{-h_{\pm}}, \qquad z \to 1.
\end{align}
The ``in'' solution has only ingoing waves at the horizon and may be written as the linear combination 
\begin{equation}\label{eq:Rin(R_+,R_-)}
R_{\rm near}^{\rm in} = A_+ R^{\widehat{\pm}}_{\rm near} + A_- R^{\widehat{\pm}}_{\rm near},
\end{equation}
with
\begin{align}\label{eq:Apm}
    A_\pm = \frac{\Gamma(1-2h_\pm)}{\Gamma(1-h_\pm-i\he)}(-i\omega/6)^{h_\pm}.
\end{align}

\subsection{Far-region}

The far equation is Eq.~\eqref{eq:R eq} with $\omega \to 0$ fixing $z$.  For general dimensions this equation has no analytic solution.  However, in the special case of five dimensions considered here, Ren \cite{Ren:2012hg,ren2013quantum} has found a remarkable change of variables that reveals the solutions to be
\begin{align}\label{eq:Rpm far}
R^{\pm}_{\rm far} = \frac{(1-z^2)^{\nu-1/2}z^{\Delta_\pm}}{(2z^2+1)^{\nu-1/2+\Delta_\pm/2}}
\,{}_2F_1\left(\frac{\Delta_\pm-1}{2}+\nu-\frac{q\ell}{\sqrt{3}},\frac{\Delta_\pm-1}{2}
+\nu+\frac{q\ell}{\sqrt{3}};\Delta_\pm-1;\frac{3z^2}{2z^2+1}\right).
\end{align}
These are normalized to unity on the boundary,
\begin{align}
    R_{\rm far}^{\pm} \sim z^{\Delta_\pm}, \qquad z \to 0.
\end{align}
The $+$ solution of interest for Dirichlet conditions has overlap region behavior given by
\begin{equation}
R^{+}_{\rm far} \sim B_+ (1-z)^{-h_+} + B_- (1-z)^{-h_-}, \qquad z \to 1,
\end{equation}
where
\begin{equation}\label{eq:Bpm}
B_\pm=\frac{(2/3)^{-1/2\mp\nu}3^{-\Delta_+/2}\Gamma(\Delta_+-1)\Gamma(\pm 2\nu)}{\Gamma\left(\frac{\Delta_+-1}{2}\pm\nu+q\ell/\sqrt{3}\right)\Gamma\left( \frac{\Delta_+-1}{2}\pm\nu-q\ell/\sqrt{3}\right)}.
\end{equation}
It is the normalization-independent ratio $\mathcal{N}=B_+/B_-$ that will affect physical quantities.

\subsection{Matching and Wronskian}

To compute the transfer function in all limits of interest, we require $R^+$ and $R^{\rm in}$ in both regions as well as their Wronskian.  We have $R^+$ in the far-region in \eqref{eq:Rpm far}; matching to the near-region gives 
\begin{equation}\label{eq:RD near}
R^{+}_{\rm near} = B_+ R^{\widehat{+}}_{\rm near} + B_- R^{\widehat{-}}_{\rm near}.
\end{equation}
On the other hand, we have $R^{\rm in}$ in the near-region in \eqref{eq:R near  pmhat}; matching to the far-region gives 
\begin{equation}\label{eq:Rfar in}
R_{\rm far}^{\rm in} = C_+ R_{\rm far}^+ + C_- R_{\rm far}^-,
\end{equation}
where
\begin{equation}\label{eq:Cpm}
C_+ = \frac{A_+ D_--A_-D_+}{B_+D_--B_-D_+},\qquad C_-=C_+\vert_{B\leftrightarrow D},
\end{equation}
with
\begin{equation}\label{eq:Dpm}
D_\pm = B_{\pm}\vert_{\Delta_+\to\Delta_-}.
\end{equation}
The Wronskian is obtained from Eq.~(13.14.25) of \cite{NIST:DLMF},
\begin{equation}\label{eq:Wronk}
\mathcal W =12A_-B_-\left(\mathcal S-\mathcal N\right)(2h_+-1),
\end{equation}
where
\begin{equation}\label{eq:S and N}
\mathcal S:=\frac{A_+}{A_-},\qquad \mathcal N:=\frac{B_+}{B_-}.
\end{equation}
Note that $\mathcal S$ is (up to normalization) just the IR CFT retarded propagator defined in the Son/Starinets prescription \cite{Son:2002sd} and that $12\nu(B_+D_--B_-D_+)=\Delta_+-2$.
 Explicitly,
\begin{equation}\label{eq:Dam son}
   \mathcal S = \mathcal{G}(-2i\omega)^{2h_+-1}, \qquad \mathcal{G}:=12^{1-2h_+} \frac{\G{1-2h_+}\G{h_+-i\he}}{\G{1-2h_-}\G{h_--i\he}}.
\end{equation}

\subsection{Parameter range with no poles}\label{sec:pole stuff}

Poles in the transfer function at frequencies $\omega_*(k)$ are excitations or ``modes'' of the theory, which correspond to instabilities if the imaginary part of the frequency is positive.  For the scalar theory we consider, it has been shown that poles for $\omega \ll 1$ always correspond to instabilities \cite{Faulkner:2009wj,Iqbal:2010eh}.  We will restrict to the parameter range where there are no poles at all for $\sigma=0$ and $\omega \ll 1$ \cite{Denef:2009tp}, which can be determined from Ren's analytical results \cite{Ren:2012hg} as follows. 

The pole condition is  $\mathcal{W}=0$,\footnote{Occasionally these can be ``false poles'', canceled by other features of the transfer function.  We have checked that in our case the remaining factors in the transfer function are all finite.} which from \eqref{eq:Wronk} is equivalently\footnote{The validity of this simplified form of the pole condition is contingent on $A_-, B_-,$ and $\mathcal{G}$ taking finite non-zero values. This the condition holds for a mode parameter set $\{k,q,m,\ell\}$ of full measure.}
\begin{align}\label{poley}
     (-2i\omega)^{2\nu} = \mathcal{N} / \mathcal{G}, %\quad A_\pm \sim \omega^{h_\pm}, \pd_\omega B_\pm=0.
\end{align}
where the RHS is independent of $\omega$.  Plugging in for the effective mass and charge, we have 
\begin{align}
    \nu(k) = \frac{1}{\sqrt{12}} \sqrt{3 + k^2+\ell^2 (m^2-2q^2)}.
\end{align}
If there is a range of $k$ such that $\nu$ is imaginary, it follows from \eqref{poley} (using the detailed form of $\mathcal{N}$ and $\mathcal{G}$) that there are always an infinite number of unstable poles.\footnote{The condition $\nu^2\geq0$ is the $\adst$ Breitenlohner-Freedman (BF) bound; this kind of instability is associated with violation of the effective ($k$-dependent) near-horizon BF bound.  Note, however, that violation of the near-horizon BF bound does not always entail instability; in fact, stable violation occurs for the Kerr spacetime and gives rise to observable power-law tails of gravitational waves \cite{Gralla:2016qfw,Gralla:2017lto,Compere:2017hsi}.}  We therefore restrict to the parameter range where $\nu$ can never become imaginary,
\begin{align}\label{BFrange}
\ell^2(m^2-2q^2) \geq -3.
\end{align}
This is equivalent to the conditions\footnote{The first condition in \eqref{nopoles} is the range where Dirichlet boundary conditions are required for decoupled $\adst$ dynamics.  However, in the present problem the effective boundary conditions for $\adst$ are not Dirichlet, but instead mixed conditions set by the ratio $\mathcal{N}$ determining the choice of boundary conditions for $\adsf$.}
\begin{align}\label{nopoles}
    m^2\ell^2 > -3, \qquad q^2\ell^2 \leq \frac{\ell^2 m^2+3}{2}.
\end{align}
There is always a range of $k$ such that $\nu$ is real, in which case $\omega^{2\nu}$ is parametrically small and poles occur only near zeros of $\mathcal{N}/\mathcal{G}$.  Noting \eqref{eq:Bpm}, such zeroes can arise only at poles of the gamma functions $ \Gamma\left( (\Delta_+-1)/2+\nu\pm q\ell/\sqrt{3} \right)$, but it is easy to show that the argument is always positive given \eqref{BFrange}.  %Given \eqref{BFrange}, the condition of no such poles is $(\Delta_+-1)/2 > |q|\ell/\sqrt{3}$, which is always satisfied.  
Thus Eqs.~\eqref{nopoles} are the complete conditions for the absence of poles, and in particular guarantee that
\begin{align}\label{eq:nu pos no poles}
    \nu>0 \qquad \textrm{(no poles for $\sigma=0$, $\omega \ll 1$)}.
\end{align}

\subsection{Near-region tail}\label{sec:horizonTail}

Taking $z$ in the near-zone and $z'$ in the far-zone defines the ``near-far'' transfer function by
\begin{equation}
    g^{\rm nf}(z,z') = \frac{R_{\rm near}^{\rm in}(z)R_{\rm far}^+(z')}{\mathcal{W}}.
\end{equation}
To compute the inverse transform we change to horizon-adapted coordinates and gauge,
\begin{align}\label{eq:GnearBpd}
    G_{\pd \rm B}^{\rm near}(v,x,z) = \frac{1}{\ell^{3/2}} \int \frac{d\omega d^3 k}{(2\pi)^4}\, e^{-i \omega v+i\vec{k}\cdot\vec{x}} \exp{\left(-\frac{i\omega}{12(1-z)}+i\he\ln(1-z)\right)} \frac{R_{\rm near}^{\rm in}(z)}{\mathcal{W}}.
\end{align}
In the above we have invoked the gauge change given in \eqref{eq:ingoing defs} and also that $e^{-i\omega t}=e^{-i\omega (v-r_*)}$ for which the quantity $r_*=-\int dz/f \sim \tfrac{1}{12}(z-1)^{-1}+ \tfrac{7}{36}\ln(1-z)$ as $z\to1$.  The gauge and coordinate transformation are both trivial on the boundary.  We have also taken the $z'$ point to the boundary using the definition \eqref{eq:G Bpd def}.  
As the frequency-domain near-zone approximation was valid at $\omega \ll 1$, the inverse transform $G_{\pd \rm B}^{\rm near}(v)$ \eqref{eq:GnearBpd} is valid at late times $v \gg 1$.\footnote{Here we assume that all other non-analytic points of the transfer function have negative imaginary parts, so that $\omega=0$ dominates at late times.  Note that only the leading non-analytic behavior near $\omega=0$ will contribute.}    That is, Eq.~\eqref{eq:GnearBpd} provides the late-time approximation to $G_{\pd \rm B}$ when the bulk point is near the horizon.

We restrict to a parameter range (Sec.~\ref{sec:pole stuff}) where $\nu$ is always real and positive.  In this case $\mathcal{S} \ll \mathcal{N}$ and $\mathcal{S}$ can be dropped from the Wronskian $\mathcal{W}$ \eqref{eq:Wronk}.  The frequency integral can then be done using Eq.~(9) in Sec.~5.20 of Ref.~\cite{bateman1954tables}, giving\footnote{The inverse transform may be understood either in the sense of Fourier or Laplace.  Since we consider the retarded Green function, which vanishes for $v - v' < 0$, the transforms differ by at most a distribution at $v=v'$, which does not contribute at late times.}
\begin{align}\label{eq:HasAretakis}
    G_{\pd \rm B}^{\rm near} = \frac{1}{\ell^{3/2}} \int \frac{d^3 k}{(2\pi)^3}\, \frac{-\Gamma(h_+-i\he)}{2\Gamma(2h_+)\Gamma(h_--i\he)B_+} (6v)^{-h_+-i\he}(1+6(1-z)v)^{-h_++i\he}.
\end{align}
The typical Aretakis behavior has now emerged at the level of the integrand, as it must on general grounds \cite{Gralla:2018xzo}.  To confirm that it remains present in the Green function requires some analysis of the integral.\footnote{A useful example to keep in mind is the extremal planar BTZ black hole (i.e., extremal BTZ unwrapped along the azimuthal direction), which is just a patch of $\ads{3}$.  The instability will be present for each momentum mode (by the general analysis of \cite{Gralla:2018xzo}), but must disappear when the integral is done because $\ads{3}$ is (linearly) stable.}  Using the fact that the $\vec{k}$-dependence is through $k^2= \vec{k}\cdot\vec{k}$ alone, we may write this in terms of $x=|\vec{x}\cdot \vec{x}|^{\frac{1}{2}}$ as
\begin{align}
G_{\pd \rm B}^{\rm near} = -\frac{1}{(2\pi)^2\,x} \frac{1}{\ell^{3/2}} \frac{(6v)^{-i\he}(1+6(1-z)v)^{i\hat e}}{(2/3)^{-1/2}\Gamma(\Delta_+-1)3^{-\Delta_+/2}}\times I,
\end{align}
where the integral $I$ is given by
\begin{align}\label{eq:I pd}
 I & = \int_0^\infty \frac{\sin(kx)}{k } \mathcal F(k^2)  dk  \\
 & = \frac{1}{2i} \int_{-\infty}^\infty \frac{e^{ikx}}{k} \mathcal F(k^2)dk \label{contour}
 \end{align}
 with 
 \begin{align}\label{eq:curly F}
\mathcal F (k^2) &:=  k^2  \frac{\Gamma(h_+-i\he)(6v)^{-h_+}(1+6(1-z)v)^{-h_+}}{\Gamma(2h_+)\Gamma(h_--i\he) (2/3)^{- \nu}\Gamma( 2\nu)}
\Gamma\left(\frac{\Delta_+-1}{2}+\nu+\frac{q\ell}{\sqrt{3}}\right) \Gamma\left(\frac{\Delta_+-1}{2}+\nu-\frac{q\ell}{\sqrt{3}}\right).
 \end{align}
\begin{figure}
 \centering
    \includegraphics[width=.6\textwidth]{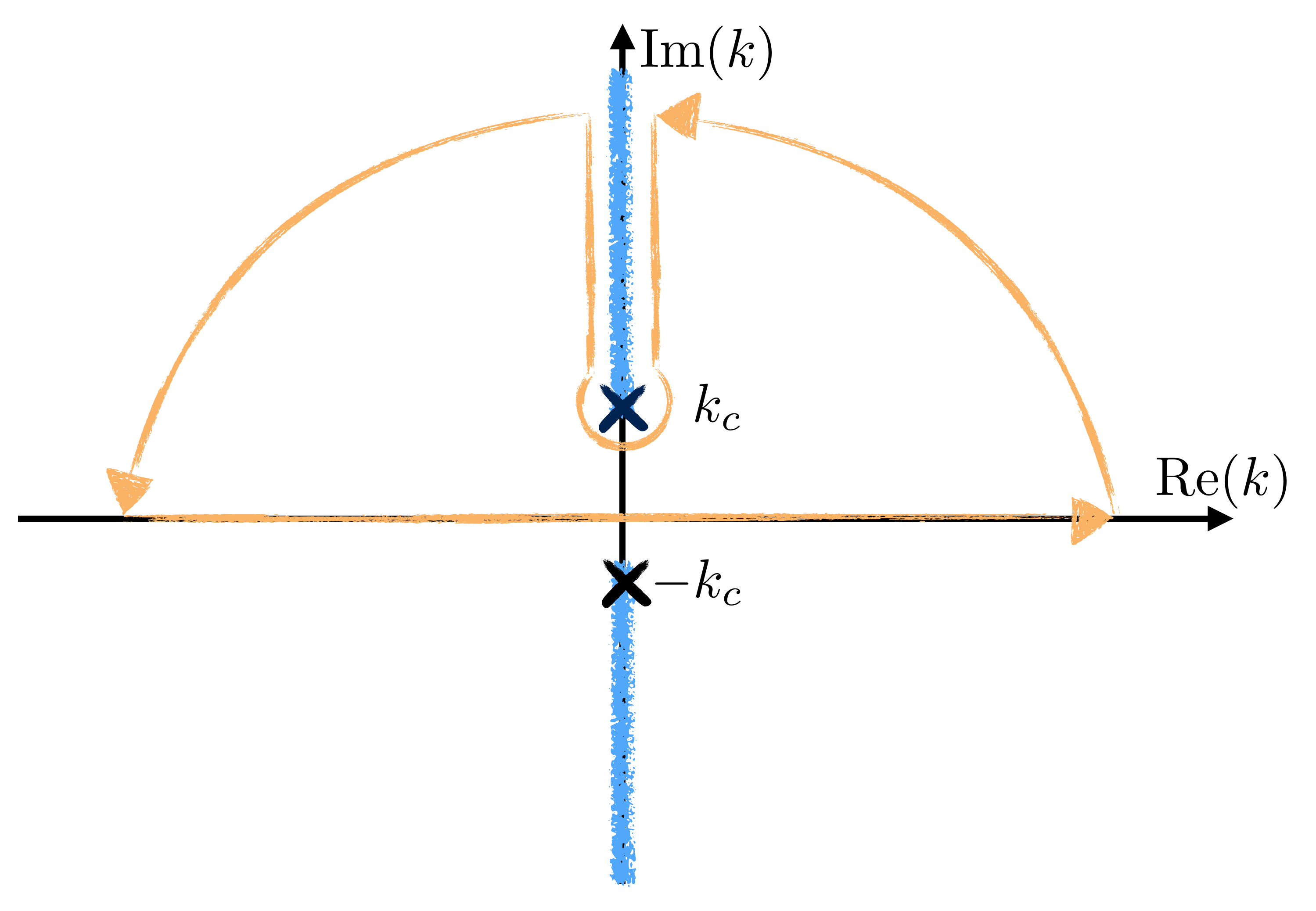}
    \caption{Illustration of the integral contour utilized in the evaluation of \eqref{eq:HasAretakis}. The branch cuts extending from the purely imaginary branch points at $\pm k_c$ where $\nu=0$ are indicated in blue.}
    \label{fig:cont}
\end{figure} 
 We compute the integral by deforming the contour in \eqref{contour} as in Fig~\ref{fig:cont}. The integrand is holomorphic everywhere except for branch points at the critical momentum values 
 \begin{align} \label{eq:xihat}
 k=\pm i/\xih, \qquad \xih = \left(3+\ell^2m^2-2\ell^2q^2\right)^{-1/2}, 
 \end{align} 
 at which $\nu=0$ and $h_\pm=1/2$.  We regard $\hat{\xi}$ as a dimensionless correlation length, following \cite{Iqbal:2011in}.
 The branch cuts and integration contour are shown in Fig.~\ref{fig:cont}.  The arc and keyhole contributions vanish,\footnote{Parametrizing the arc in polar form $k=R e^{i \theta}$, we see that $
\abs{\mathcal F_f} \sim N R^2 v^{-1/2-R}(1+6(1-z)v)^{-1/2-R}$ as $R \to \infty$
for some constant $N$, satisfying the conditions of Jordan's lemma.  (The ratio of $\Gamma$ functions in \eqref{eq:curly F} is order unity.)  Similarly, by writing $\delta e^{i\theta} = k - i/\xih$, the keyhole contribution from the circle of radius $\delta$ about $k=i/\xih$ (where $\nu=0$) is seen to be suppressed as $\delta \to 0$ on account of the element  $dk=\delta e^{i\theta}id\theta$. As the line element multiplies a function of $\delta$ with a smooth limit to $\delta=0$, if follows that this portion of the integral vanishes linearly in the key-hole radius $\delta$.} leaving just the branch cut difference along $\rho=i k$,  
\begin{equation}\label{eq:I pd result}
I=\frac{1}{2i}\int_{1/\xih}^\infty\frac{e^{-\rho x}}{\rho}\left(\mathcal F(-\rho^2)-\mathcal F(-\rho^2)\vert_{\nu\to-\nu}\right) d\rho.
\end{equation}
This integral is now in a form convenient for numerical analysis.  We may proceed analytically by noting that, on account of the exponential factor $e^{-x/\xih}$, for large $x$ the main contribution comes from near $1/\xih$,
\begin{equation}\label{eq:Fpd near kc}
    \mathcal F(\rho^2)\sim -2i\sqrt{2} \xih^{-5/2} 
  (6v)^{-1/2}(1+6(1-z)v)^{-1/2}\Gamma\left(\frac{\Dp-1}{2}+\frac{q\ell}{\sqrt{3}}\right)\Gamma\left(\frac{\Dp-1}{2}-\frac{q\ell}{\sqrt{3}}\right) 
  \sqrt{\rho-1/\xih}, \qquad \rho\to 1/\xih.
\end{equation}
Noting that this expression flips sign under $\nu \to -\nu$, 
the integral \eqref{eq:I pd result} at large values of $x$ is given by\footnote{Here we have made use of the gamma function type integral $\int_0^\infty e^{-az}\sqrt{z} dz=\sqrt{\pi}/(2a^{3/2})$ for $a>0$.}
\begin{equation}\label{eq:I pd large x}
    I \sim -\sqrt{2\pi}\, \Gamma\!\left(\frac{\Dp-1}{2}+\frac{q\ell}{\sqrt{3}}\right)\Gamma\!\left(\frac{\Dp-1}{2}-\frac{q\ell}{\sqrt{3}}\right)\, \left(\frac{1}{x\xih}\right)^{3/2}
  (6v)^{-1/2}(1+6(1-z)v)^{-1/2} e^{-x/\xih},\qquad\! x\to\infty.
\end{equation}

The late-time, large-$x$, near-zone bulk-boundary Green function is therefore
\begin{align}\label{eq:G pd large x}
    G^{\rm near}_{\pd\rm B} \sim 
   \frac{\mathcal C}{\ell^{3/2}x}\,\left(\frac{1}{x\xih}\right)^{3/2}
  (6v)^{-1/2-i\he}(1+6(1-z)v)^{-1/2+i\he} e^{-x/\xih}, \qquad x\to\infty, 
\end{align}
where 
\begin{equation}\label{eq:C pd}
  \mathcal C := \frac{3^{(\Dp-1)/2}}{2\pi^{3/2}} \frac{
    \Gamma\left(\frac{\Dp-1}{2}+\frac{q\ell}{\sqrt{3}}\right)\Gamma\left(\frac{\Dp-1}{2}-\frac{q\ell}{\sqrt{3}}\right)}{\Gamma(\Delta_+-1)}.   
\end{equation}
Notice that the $v^{-1/2}$ decay at large $x$ is equal to the decay of the mode at the critical momentum $k_c=i/\xih$.  Restoring dimensional factors gives Eq.~\eqref{money} of the introduction.  

Equation~\eqref{eq:G pd large x} shows explicitly that the Aretakis instability exists for a source (or initial data) localized near the boundary and evaluation point at large transverse distance $x$ from the support of the source.   
However, these assumptions are only for convenience; the instability persists for initial data extending into the bulk and for any point $x$ on the horizon.  Allowing the data to extend into the bulk amounts to carrying around the factor $R^+_{\rm far}(z')$, which is $\omega$-independent and hence does not affect the analysis.\footnote{The analysis may require modification if the data extends all the way to the horizon, where $R^+$ is not regular.}  Including finite values of $x$ means we must deal with the full integral over $k$-space.  However, in the form \eqref{eq:I pd result} the integral is absolutely convergent, so we may bring $z$-derivatives inside, where they pull down powers of $v$.  The contribution from $\rho \sim 1/\xi$ will grow after only a single $z$-derivative [see Eq.~\eqref{eq:Fpd near kc}], so we expect the integral to grow in time.  We have verified this fact numerically: the instability indeed persists at the single $z$-derivative level at any $x$.

\subsection{Far-region tail} \label{far tail stuff}

For the inverse transform in the far-region we work in the original coordinates and gauge. The far-far transfer function is given by
\begin{equation}\label{eq:gff}
    g^{\rm ff}:= \frac{\left(S(\omega) R^+_{\rm far}(z_>)+R^-_{\rm far}(z_>)\right)R^+_{\rm far}(z_<)}{2\Delta_+-4},
\end{equation}
where
\begin{equation}\label{eq:S}
    S(\omega):=\frac{C_+(\omega)}{C_-(\omega)} =-\frac{D_-\mathcal G (-2i\omega)^{2h_+-1}-D_+}{B_-\mathcal G (-2i\omega)^{2h_+-1}-B_+}.
\end{equation}
 The quantity $S$ is the small-$\omega$ approximation to the dual two-point function in the Son/Starinets prescription; Equation~\eqref{eq:S} may be compared with Eq.~(54) of Ref.~\cite{Faulkner:2009wj}.  Recall also from Eq.~\eqref{eq:Dam son} that $\mathcal{S}=\mathcal{G}(-2 i \omega)^{2h_+-1}$ is the ``IR two-point function''.  In Eqs.~\eqref{eq:gff} and \eqref{eq:S}, all of the $\omega$-dependence is explicit.
 
Notice that $g^{\rm ff}$ takes the form
\begin{equation}
    g^{\rm ff}\sim \frac{D_- \mathcal G}{(2\Delta_+-4)B_+} (-2 i\omega)^{2h_+-1} R^+_{\rm far}(z_>)R^+_{\rm far}(z_<) + \textrm{(terms smooth in $\omega$)}, \qquad \omega \to 0.
\end{equation}
The smooth terms do not contribute at late times.  We define the ``far'' propagator in the time domain to include only the late-time behavior,
\begin{align}\label{eq:GfarBpd}
    G^{\rm far}_{\pd\rm B}(\tau,x,z) = \frac{1}{\ell^{3/2}}\int \frac{d\omega d^3 k}{(2\pi)^4}\, e^{-i \omega \tau+i\vec{k}\cdot\vec{x}} \frac{D_- \mathcal G}{(2\Delta_+-4)B_+} (-2 i\omega)^{2h_+-1} R^+_{\rm far}(z).
\end{align}
For convenience we have also taken $z'$ to the boundary, using the definition \eqref{eq:G Bpd def}.  Equation~\eqref{eq:GfarBpd} provides the late-time approximation to $G_{\pd \rm B}$ when the bulk point is away from the horizon.  

The inverse transform in time is trivial,
\begin{align}\label{eq:G far B pd tail}
    G^{\rm far}_{\pd\rm B}(\tau,x,z) = \frac{1}{\ell^{3/2}4(\Delta_+-2)}\int \frac{d^3 k}{(2\pi)^3}\, e^{i\vec{k}\cdot\vec{x}} \frac{D_- \mathcal G}{B_+\Gamma(1-2h_+)} (\tau/2)^{-2h_+} R^+_{\rm far}(z),
\end{align}
but as in the near-zone case a complicated $k$ integral remains.  Paralleling the discussion of Sec.~\ref{sec:horizonTail}, we write
\begin{align}
G_{\pd \rm B}^{\rm far} = \frac{\G{\Dm-1}3^{\frac12(\Dp-\Dm)}}{(2\pi)^2x\ell^{3/2}\G{\Dp-1} (\Delta_+-2)} \times I_f,
\end{align}
where the integral $I_f$ is given by
\begin{align}\label{eq:I f}
 I_f & = \int_0^\infty \frac{\sin(kx)}{k } \mathcal F_f(k^2)  dk  \\
 & = \frac{1}{2i} \int_{-\infty}^\infty \frac{e^{ikx}}{k} \mathcal F_f(k^2)dk \label{contour}
 \end{align}
 with 
 \begin{align}\label{eq:curly F far}
\mathcal F_f (k^2) &:=   R_{\rm far}^+(z) k^2 \frac{9^{-2\nu}\G{h_+-i\he}\G{-2\nu}\G{\frac{1}{2}(\Dp-1)+\nu+q\ell/\sqrt{3}}\G{\frac12(\Dp-1)+\nu-q\ell/\sqrt{3}}}{\G{h_--i\he}\G{2\nu}^2\G{\frac{1}{2}(\Dm-1)-\nu+q\ell\sqrt{3} }\G{\frac{1}{2}(\Dm-1)-\nu-q\ell/\sqrt{3}}\tau^{2h}}.
 \end{align}
  We use the same integration contour as above (Fig.~\ref{fig:cont}).   The arc and keyhole contributions again vanish,\footnote{To show the applicability of Jordan's Lemma, we again parametrize the arc in polar form and use Stirling's approximation to derive that $\abs{\mathcal F_{f}} \simeq  N R^2 t^{-2R}9^{-2R}$ as  $\quad R \to \infty$.   Similarly, the keyhole contribution (where $\nu=0$) is easily seen to be suppressed on account of the element  $dk = \delta e^{i\theta} i d\theta$  for a radius-$\delta$ circle about $k_c$  ($\delta e^{i\theta} = k - k_c$) as $\delta \to 0$ and the presence of the factor $\Gamma(-2\nu)/\Gamma(2\nu)^2\propto \sqrt{\delta}$. These $\delta$-suppressed terms  multiply a smooth function of $\delta$  and therefore the key-hole integral vanishes as $\delta^{3/2}$.} leaving just the branch cut difference 
 \begin{equation}
I_{f} = \frac{1}{2i}
 \int_{1/\xih}^\infty
\frac{e^{-\rho x}}{\rho} \left(\mathcal F_{f}(-\rho^2)- \mathcal F_{f}(-\rho^2)\vert_{\nu\to-\nu}\right) d\rho.
\end{equation}
in terms of $\rho=ik$.  The integral is now in a form convenient for numerical analysis.  As before, we may proceed analytically by noting that the large-$x$ contribution is dominated by $\nu \approx 0$.  The end result is
\begin{equation}\label{eq:G pdpd tail}
    G^{\rm far}_{\pd\rm B}\sim \frac{\mathcal C_{f}}{\ell^{3/2}x}\left(\frac{1}{\xih x}\right)^{3/2}\tau^{-1} e^{-x/\xih} R^+_{\rm far}(z)|_{h=1/2},\qquad x\to\infty,
\end{equation}
where 
\begin{equation}\label{eq:C pdpd}
    \mathcal C_{f}=\frac{1}{2\sqrt{2}\pi^{3/2}}\frac{3^{\frac12(\Dp-\Dm)}
    \G{\Dm-1}
    \G{\frac{\Dp-1}{2}+\frac{q\ell}{\sqrt{3}}} \G{\frac{\Dp-1}{2}-\frac{q\ell}{\sqrt{3}}}}{(\Delta_+-2)
    \G{\Dp-1} 
    \G{\frac{\Dm-1}{2}+\frac{q\ell}{\sqrt{3}}} 
    \G{\frac{\Dm-1}{2}-\frac{q\ell}{\sqrt{3}}}}.
\end{equation}

\section{Near-extremal case}\label{sec:less extreme}

We now consider the near-extremal case $\sigma \ll 1$, where the Aretakis behavior occurs transiently.  Following \cite{Gralla:2016sxp,Zimmerman:2016qtn,Compere:2017hsi}, we will be able to analytically exhibit the transition from power-law growth to exponential decay by performing a sum over quasinormal modes.  The relevant modes form an infinite chain descending from near the origin $\omega=0$ with imaginary parts separated by $\sigma$. Though each mode decays exponentially, as $\sigma\to0$, all values of $n$ become important and the modes sum coherently. The coherent sum reveals a power-law decay at times $t\ll 1/\sigma$ with transient semi-local critical features which become permanent in the extremal limit.  
This can be interpreted as a line of near-extremal poles coalescing into an extremal branch cut.

The near-extremal calculation is also done with matched asymptotic expansions, using the same regimes \eqref{eq:MAE}.  The analysis of the far and overlap regions is identical to the extremal case; the only new behavior emerges at $1-z\sim \sigma$ in the near-region.  The near-region radial equation is now given by $\sigma \to 0$ with $\omega \sim 1-z \sim \sigma$,
\begin{align}\label{eq:near near radial eq} 
\left((1-z)\left(1-z+\sigma\right)\Ru_{\rm near}'\right)'
+\Bigg(\frac{(\omega/12+\he(1-z))^2}{(1-z)(1-z+\sigma)}-\hat{m}^2\Bigg)\Ru_{\rm near}=0,
\end{align}
where $\he$ and $\hat{m}^2$ are the effective $\ads{2}$ charge and mass parameters given in \eqref{eq:ads2 charge and mass}.
Here and below, we use an underline to denote near-extremal quantities that reduce to corresponding non-underlined quantities as $\sigma \to 0$.

The solution with pure ingoing waves at the horizon is \cite{Faulkner:2011tm}
\begin{equation}\label{eq:Rin near T}
\Ru^{\rm in}_{\rm near}=\left(\frac{1-z}{\sigma}\right)^{-\frac{i\omega}{12\sigma}} \left(1+\frac{1-z}{\sigma}\right)^{\frac{i\omega}{12\sigma}-i\hat e}{}_2F_1\left(h_+-i\he,1-h_+-i\hat e;1-\frac{i\omega}{6\sigma};-\frac{1-z}{\sigma}\right).
\end{equation}
  The confluence identity $W_{a,b}(z) = \lim_{c\to\infty} {}_2 F_{1}\left(b-a-1/2,1/2-b-a;c;1-c/z\right) e^{-z/2} z^{a}$ may be used to see that \eqref{eq:Rin near T} properly reduces to the last equation in \eqref{eq:R near  pmhat}.

The overlap region behavior is
\begin{align}\label{eq:Rin overlap T}
    \underline R^{\rm in}_{\rm near} \sim \Au_+ (1-z)^{-\hp} + \Au_- (1-z)^{-\hm}, \qquad z \to 1,
\end{align}
where 
 \begin{equation}\label{eq:Apm T}
     \Au_\pm = \frac{\Gamma\Big(1-i\omega/(6\sigma)\Big)\Gamma(1-2h_\pm)}
    {\Gamma(1-h_\pm-i\he)\Gamma(1-h_\pm-i\omega/(6\sigma)+i\he)}\sigma^{h_\pm-i\he}.
\end{equation}
We also introduce the ratio
\begin{align}
    \Su=\frac{\Au_+}{\Au_-},
\end{align}
which can be obtained from $\mathcal S$ in \eqref{eq:S and N} by noting that
\begin{equation}\label{eq:Su to S}
   \mathcal  S\to \underline{\mathcal S} \quad \text{   under   } \quad  (-2i\omega)^{2h_+-1}\to \frac{\Gamma(h_+-i\omega/(6\sigma)+i\he)}{\Gamma(h_--i\omega/(6\sigma)+i\he)}\sigma^{h_+-h_-}.
\end{equation}

We again consider the parameter regime (Sec.~\ref{sec:pole stuff}) where $\nu > 0$.  This guarantees that there are no small-$\omega$ poles at $\sigma=0$ (which would correspond to instabilities), but new, stable poles appear in the near-extremal case for $\omega \sim \sigma$.  The Wronskian is given by \eqref{eq:Wronk} with $A_\pm \to \Au_\pm$ ($\mathcal{S}\to\Su$) and the pole condition $\mathcal{W}=0$ becomes $\Su=\mathcal{N}$, which  can be stated explicitly as
\begin{equation}\label{eq:mommy}
    \frac{\Gamma(1-h_--i\omega/(6\sigma)+i\he)}
    {\Gamma(1-h_+-i\omega/(6\sigma)+i\he)}\sigma^{2\nu}= \text{(constant independent of $\omega$ and $\sigma$)}.
\end{equation}
In the $\nu>0$ parameter range considered here the quantity $\sigma^{2\nu}$ is parametrically small, and this equation can only be satisfied if one approaches a pole of the gamma function in the numerator as $\sigma \to 0$.  The solutions are thus
\begin{equation}\label{eq:modes}
\omega_n = 6 \sigma \left(\hat e-i(h_++n)\right)+O(\sigma^{2\nu}), \qquad n=0,1,2,\dots \,.
\end{equation}
Notice that the modes are arbitrarily long-lived in the limit $\sigma \to 0$, since $\textrm{Im}[\omega_n]\to 0$. Such modes were first studied in Refs.~\cite{Hod:2008zz,Yang:2013uba}; some common names are ``zero-damped modes'' and ``near-horizon modes''.

\subsection{Near-region ring-up}

We now provide near-extremal versions of the results of Sec.~\ref{sec:horizonTail}.  The near-region bulk-boundary retarded Green function is given by Eq.~\eqref{eq:GnearBpd} with $R^{\rm in}_{\rm near} \to \underline{R}^{\rm in}_{\rm near}$ and similar replacements in the Wronskian $\mathcal{W}$ (see text above Eq.~\eqref{eq:mommy}).  Noting that $\underline{\mathcal{S}} \ll \mathcal{N}$ can be dropped in the $\nu>0$ parameter range considered here, we find
\begin{equation}\label{eq:GnearBpd T}
    \underline{G}^{\rm near}_{\mathrm B\pd}(v,x,z)=\frac{1}{\ell^{3/2}}\int \frac{d\omega d^3k}{(2\pi)^4} e^{-i\omega v+i\vec{k}\cdot\vec{x}} \frac{1}{{12(1-2h_+)\underline{A}_-B_+}} {}_2F_1\left(h_+-i\he,1-h_+-i\hat e;1-\frac{i\omega}{6\sigma};-\frac{1-z}{\sigma}\right).
\end{equation}

In the extremal case \eqref{eq:GnearBpd}, we were able to perform the complete frequency integral analytically.  In the near-extremal case, we shall content ourselves with the portion due to quasinormal modes (poles in the complex-$\omega$ plane), which is expected to dominate at late times.  The mode spectrum was given previously in \eqref{eq:modes}; computing the residues resolves the $\omega$-integral of Eq.~\eqref{eq:GnearBpd T} into a discrete sum:
\begin{align}\label{eq:g near pd qnm n T}
\underline{G}^{\rm near}_{\mathrm B\pd}(v,x,z)=\frac{-1}{\ell^{3/2}}\int &\frac{d^3k}{(2\pi)^3} e^{i\vec{k}\cdot\vec{x}} \Bigg(
\sigma^{h_++i\he} e^{-(h_++i\he)6\sigma v} \frac{\Gamma(h_+-i\he)}{2\Gamma(2h_+)B_+} \nonumber \\ &\times\sum_{n=0}^\infty \frac{(-1)^n e^{- 6 n  \sigma v}}{n!\Gamma(h_--i\he-n)}{}_2F_1(h_+-i\he,1-h_+-i\he,1-h_+-i\he-n,-(1-z)/\sigma) \Bigg).
\end{align}
Remarkably, this sum can be performed in closed form \cite{Gralla:2016sxp}.  Using the series definition of the hypergeometric function \cite{NIST:DLMF} and commuting the order of summation, the result is seen to be 
    \begin{align}\label{eq:g near pd qnm T}
     \underline{G}^{\rm near}_{\mathrm B\pd}(v,x,z) &=\frac{-1}{\ell^{3/2}} \int \frac{d^3k}{(2\pi)^3}  e^{i\vec{k}\cdot\vec{x}}  \frac{e^{-(h_++i\he)6\sigma v}\Gamma(h_+-i\he)}{2\Gamma(2h_+)\Gamma(h_--i\he)B_+} \left(\frac{1-e^{-6 \sigma v}}{\sigma}\right)^{-h_+-i\he}
     \Big(1+\frac{1-z}{\sigma}\left(1-e^{-6 \sigma v}\right)\Big)^{-h_++i\he}.
\end{align}
Equation~\eqref{eq:g near pd qnm T} provides the late-time ($v\sim1/\sigma$) approximation to $G_{\pd \rm B}$ when the bulk point is near the horizon. Notice that \eqref{eq:g near pd qnm T} may be obtained from the extremal version, \eqref{eq:HasAretakis}, simply by sending $v \to (1 - e^{-6 \sigma v})/(6 \sigma)$ and then multiplying by $e^{-(h_++i\he)6\sigma v}$ inside the integral.  We may therefore determine the large-$x$ approximation by performing the same substitution in the large-$x$ extremal result \eqref{eq:G pd large x}, giving 
\begin{equation}\label{eq:G near pd qnm}
     \underline{G}^{\rm near}_{\mathrm B\pd}\sim 
   \frac{\mathcal C}{\ell^{3/2}x}\left(\frac{1}{x \xih}\right)^{3/2}
  e^{-(1/2+i\he)6\sigma v}
 \left(\frac{1-e^{-6 \sigma v}}{\sigma}\right)^{-1/2-i\he}\Bigg(1+(1-z)\left(\frac{1-e^{-6 \sigma v}}{\sigma}\right)\Bigg)^{-1/2+i\he}
\! e^{-x/\xih}, \qquad x\to\infty, 
\end{equation}
where $\mathcal C$ was given previously in Eq.~\eqref{eq:C pd}.  Equation~\eqref{eq:G near pd qnm} reduces to the extremal result \eqref{eq:G pd large x} as $\sigma \to 0$, which physically corresponds to early times $v \ll 1/\sigma$.  However, since we have made a late-time approximation $v \gg 1$ in deriving Eq.~\eqref{eq:G near pd qnm}, the regime where the Aretakis behavior \eqref{eq:G pd large x} emerges is in fact intermediate times $1 \ll v \ll 1/\sigma$.
\begin{figure}
 \centering
    \includegraphics[width=.7\textwidth]{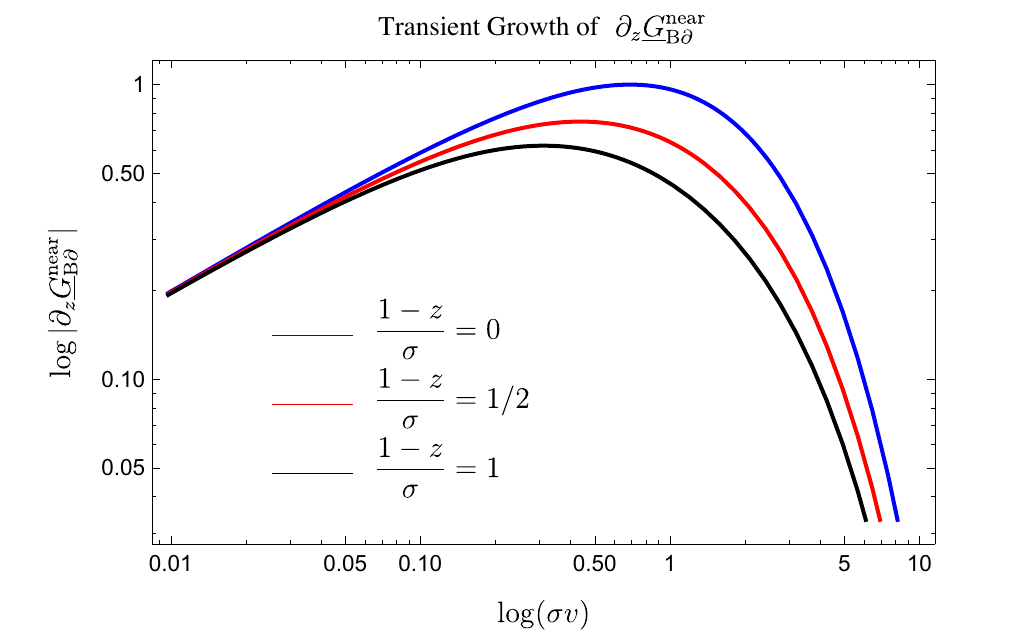}
    \caption{Radial derivative of the quasinormal mode-summed near-horizon bulk-boundary Green function $\underline{G}^{\rm near}_{\mathrm B\pd}$ \eqref{eq:g near pd qnm T} normalized by its maximum value attained on the horizon. The transient Aretakis behavior, corresponding in this case to $\sqrt{v}$ growth, occurs in the regime $v\sigma\ll1$ before the exponential decay of the lowest-lying quasinormal mode takes over.}
    \label{fig:Gnear}
\end{figure} 

\subsection{Far-region ring-down}\label{sec:ring down}

The far-far transfer function has the same form as the extremal version \eqref{eq:gff},
\begin{align}\label{eq:far far T}
 \underline{g}^{\mathrm {ff}} =\frac{\left(\underline{S}(\omega)R^+_{\rm far}(z_>)+R^-_{\rm far}(z_>)\right)R^+_{\rm far}(z_<)}{2\Delta_+-4},
\end{align}
where now
\begin{align}\label{eq:far SS T}
\underline{S} =\frac{D_- \underline{\mathcal S}-D_+}{B_-(\mathcal N-\underline{\mathcal S})},
\end{align}
which can be obtained from \eqref{eq:S} using the transformation of $(-2i\omega)^{2h_+-1}$ given in \eqref{eq:Su to S}.

 In the extremal case, we kept only the leading non-analytic behavior of $g^{\rm ff}$ as $\omega \to 0$.  We defined $G^{\rm far}_{\rm B \pd}$ to be the inverse transform of this piece only, meaning it displays only the leading power-law behavior at late times.  In the near-extremal case, $g^{\rm ff}$ is analytic at $\omega=0$ (there is no power-law tail), but a series of poles \eqref{eq:modes} appears on the negative imaginary axis.  We similarly define $G^{\rm far}_{\pd B}$ to be just the pole contribution to the inverse transform,
 \begin{align}\label{eq:qnm sum g pdpd}
   \underline{G}^{\rm far}_{\pd B}(\tau,x,z)
   &=\frac{1}{\ell^{3/2}4(\Delta_+-2)}\int \frac{d^3 k}{(2\pi)^3} e^{i\vec{k}\cdot \vec{x}}(12\sigma)^{2h_+}\frac{D_-\mathcal{G}}{B_+}e^{-(h_++i\he)6 \sigma \tau}R_{\rm far}^+(z) \sum_{n=0}^\infty \frac{(-1)^n e^{-6\sigma n\tau}}{n!\Gamma(1-2h_+-n)}\nonumber\\
   &=\frac{1}{\ell^{3/2}4(\Delta_+-2)}\int \frac{d^3 k}{(2\pi)^3} e^{i\vec{k}\cdot \vec{x}}(12\sigma)^{2h_+}\frac{D_-\mathcal{G}}{B_+\Gamma(1-2h_+)}e^{-(h_++i\he)6 \sigma \tau}(1-e^{-6 \sigma \tau})^{-2h_+}R_{\rm far}^+(z),
\end{align}
where $\mathcal G$ was given previously in  \eqref{eq:Dam son}. 

Equation~\eqref{eq:qnm sum g pdpd} provides the late-time approximation to $G_{\pd \rm B}$ when the bulk point is a finite distance outside the horizon.  Notice that it may be obtained from extremal version \eqref{eq:G far B pd tail} by sending $\tau \to (1 - e^{-6 \sigma \tau})/(6 \sigma)$ and multiplying by $e^{-(h_++i\he)6\sigma \tau}$ inside the integral.  We may therefore determine the large-$x$ approximation by performing the same substitution in the large-$x$ extremal result \eqref{eq:G pdpd tail}, giving
\begin{align}\label{eq:G pdpd tail T}
  \underline{G}^{\rm far}_{\mathrm{B}\pd}\sim \mathcal C_{f} \frac{6 \sigma}{x^4 \ell^{3/2}}\left(\frac{x}{\xih}\right)^{3/2} e^{-(1/2+i\he)6 \sigma  \tau} (1-e^{-6 \sigma \tau})^{-1} e^{-x/\xih} R^+_{\rm far}(z)\vert_{h=1/2},\qquad x\to\infty.
\end{align}
Equation~\eqref{eq:G pdpd tail T} reduces to the extremal result \eqref{eq:G pdpd tail} as $\sigma \to 0$, which physically corresponds to early times $v \ll 1/\sigma$.  However, since we have made a late-time approximation $v \gg 1$ in deriving Eq.~\eqref{eq:G pdpd tail T}, the regime where the power-law behavior \eqref{eq:G pdpd tail} emerges is in fact intermediate times $1 \ll v \ll 1/\sigma$.

\section{What is the CFT dual?}\label{sec:interactions}

Our main motivation in this paper has been to explore the holographic implications of the Aretakis instability.  We have now demonstrated that the instability persists in a setting of relevance to holography, but we have not identified a dual description in terms of field theory degrees of freedom.  Finding a precise dual description seems quite challenging, as the physical effects of the instability involve an infalling observer, who measures large stress-energies as she crosses the horizon.  However, the Aretakis instability provides a good target within the general goal of finding the dual description of deep-bulk phenomena, since we would naturally expect it to correspond to something large in the field theory.  If we could identify a field-theory quantity that grows with time, this could provide a first tentative entry in the long-sought deep-bulk regime of the holographic dictionary.

To address this question we must go beyond the free scalar theory.  There is no boundary growth in the free theory---the field and all its derivatives decay everywhere outside the horizon---and a free scalar is not expected to enjoy a complete duality, anyway.  However, we may hope that by adding some interactions, we can capture enough features of a full duality to see whatever is dual to the Aretakis growth.  In particular, we might imagine a scattering experiment that probes the near-horizon growth and brings the information back to the boundary.  This could plausibly manifest as growth of boundary correlators, or at least as some other distinctive late-time behavior.  

Unfortunately, this story is at best only partially true.  While the Aretakis instability does seem to influence late-time behavior of boundary correlators, we find no sign of any kind of growth.  In fact, the instability seems to enforce a boundary conformal symmetry, which in turn enables a simple argument suggesting that \textit{all} $n$-point correlators decay in time at \textit{any} order in the $1/N$ expansion.  Thus there is no obvious CFT dual description of the Aretakis growth in the standard perturbative holographic dictionary.  However, our argument is only meant to be suggestive, and leaves out many details.  It therefore remains possible that some kind of anomaly will arise in a more careful calculation.

\subsection{Temporal conformal symmetry}\label{sec:TCS}

To make the argument we return to the notation of the introduction and again set the charge $\hat{e}$ of the field to zero.  Recall that the dual two-point function was seen to decay linearly in time in a certain regime \eqref{regime} of large spatial and temporal separation.  We noted that this behavior is captured by the scaling behavior
\begin{align}\label{2ptsym}
    \langle \OO_1 \OO_2 \rangle \to \lambda \langle \OO_1 \OO_2 \rangle \textrm{ under } t_i \to \lambda^{-1}  t_i.
\end{align}
Here $\OO_n=\OO(t_n,\vec{y}_n)$ means the dual operator evaluated at the $4D$ boundary point $n$.  The two-point function can only depend on the time difference $\delta t_{12}=t_2-t_1$, but for simplicity we rescale both points separately.  %We refer to $t_i \to \lambda^{-1} t_i$ as time dilation.  

The scaling \eqref{2ptsym} implies the linear decay $\langle \OO_1 \OO_2 \rangle \sim 1/\delta t_{21}$ of the correlator, which is of course how we arrived at it in the first place.  But we may see the decay directly by regarding $t_i \to \lambda^{-1} t_i$ as defining a flow to late times as $\lambda \to 0$; Eq.~\eqref{2ptsym} states that $\langle \OO_1 \OO_2 \rangle$ vanishes linearly in $\lambda$.  (For simplicity we set $T=0$ so that the flow may proceed to arbitrarily late times without breaking out of the regime of validity \eqref{regime}.  This is really just shorthand for working in the regime \eqref{regime}.) This approach will generalize to higher-point correlators, whose precise form will be more difficult to construct.  For example, we will argue below that the tree-level (leading in $1/N$) four-point correlator dual to $\Phi^4$ theory satisfies 
\begin{align}\label{4ptsym}
    \langle \OO_1 \OO_2 \OO_3 \OO_4 \rangle \to \lambda^2 \langle \OO_1 \OO_2 \OO_3 \OO_4 \rangle \textrm{ under } t_i \to \lambda^{-1}  t_i
\end{align}
when all space and time differences $\delta t_{ij}=t_j-t_i$ and $\delta y_{ij}=|\vec{y}_j-\vec{y}_i|$ (with $i \neq j$) are in the large-separation regime \eqref{regime}.  This provides a sense in which the correlator decays in time: $\langle \OO_1 \OO_2 \OO_3 \OO_4 \rangle$ vanishes quadratically in $\lambda$ under the flow $t_i \to \lambda^{-1}  t_i$.  While Eq.~\eqref{4ptsym} is consistent with simple power laws such as $\langle \OO_1 \OO_2 \OO_3 \OO_4 \rangle \sim 1/(\delta t_{12} \delta t_{34})$, it permits far more general behavior.  It would be very interesting to compute the precise form of $\langle \OO_1 \OO_2 \OO_3 \OO_4 \rangle$, but for the present we shall content ourselves with arguing for the scaling \eqref{4ptsym}.

We can view the properties \eqref{2ptsym} and \eqref{4ptsym} as a symmetry under the rescalings
\begin{align}\label{tconf}
    %\textrm{temporal conformal transformation: } 
    \qquad t \to \lambda^{-1} t, \ \OO \to \lambda^{-1/2} \OO,
\end{align}
which we will refer to as a \textit{temporal conformal transformation}.  Equations~\eqref{2ptsym} and \eqref{4ptsym} state that the leading two- and four-point functions enjoy temporal conformal symmetry in the large-separation regime.  We will argue that, in fact, all $n$-point correlators continue to enjoy the emergent symmetry at any order in $1/N$ and in any interacting theory.  This provides a sense in which all holographic correlators decay in time.

%Even if the symmetry does becomes anamalous because of a subtlety invisible to the formal argument, it would be hard to attribute this anomaly to the Aretakis instability, which seems to play a central role in \textit{enforcing} that the symmetry be unbroken.

\subsection{Four-point function in $\Lambda \Phi^4$ theory}

Let us imagine adding a $\Lambda \Phi^4$ interaction to our scalar theory and computing a four-point function.  We will begin with a connected retarded four-point function at tree-level (leading order in ``$1/N$''), given by the integral
\begin{align}\label{4point}
    \langle \OO_1 \OO_2 \OO_3 \OO_4 \rangle & = \Lambda \int \sqrt{-g} dV dR d^3 \vec{y}\, G_{\pd \rm B}^{1} G_{\pd\rm B}^{2} G_{\rm B \pd}^{3} G_{\rm B \pd}^{4}. 
\end{align}
%Here $\OO_n=\OO(V_n,\vec{y}_n)$ means the dual operator evaluated at the $4D$ boundary point $n$, 
Here $G_{\pd\rm B}^n=G_{\pd\rm B}(V_n, \vec{y}_n; V,r,\vec{y})$ is the boundary-to-bulk retarded propagator from the $4D$ boundary point $n$ to the $5D$ spacetime integration point.  (Similarly, $G_{\rm B \pd}^n$ means retarded propagation from bulk integration point to boundary spacetime point, a propagator we have not studied in the text.)  We do not make precise the meaning of $\langle \OO_1 \OO_2 \OO_3 \OO_4 \rangle$, simply regarding \eqref{4point} as its tree-level definition from holography (we refer to the boundary-limit dictionary \cite{Banks:1998dd,Harlow:2011ke}).  We work formally, without worrying about convergence.

By symmetry, $\langle \OO_1 \OO_2 \OO_3 \OO_4 \rangle$ can only depend on time and space differences $\delta t_{12}=t_2-t_1$ and $\delta y_{12}=|\vec{y}_2-\vec{y}_1|$ (and all other combinations of spacetime points).  %For simplicity we will set $T=0$ and assume that all time and space points are large, ($V_n,\vec{y}_n \gg 1/\mu$), 
We will consider the case where all such differences are in the regime \eqref{regime}, which we refer to as the large-separation regime.  The range of integration of the bulk point is unrestricted, but based on the physical picture of propagation (where $\langle \OO_1 \OO_2 \OO_3 \OO_4 \rangle$ involves propagation to an interaction point and then back away from it), we expect the integral to be dominated by large separations in each leg, i.e., large values of  $V-V_n$ and $|\vec{y}-\vec{y}_n|$.  We will therefore replace the propagators with their large-separation approximations \eqref{money} in Eq.~\eqref{4point}.\footnote{This substitution will actually cause the integral to diverge, since a late-time power-law decay $1/t^p$ looks like a divergence at early times.  So, strictly speaking we impose a cutoff that restricts the integral to the regime of validity of the late-time approximation.  A more rigorous argument would contain a proof that the neglected portion is in fact subleading as boundary space and time separations become large.} This will entail splitting up the radial integral into near and far regions.  We will see that the near-region dominates and provides the expected scaling \eqref{4ptsym}.

Now we send $t_i \to \lambda^{-1} t_i$ in the four-point function \eqref{4point}.  Noting that the boundary time $t$ agrees with the bulk coordinate $V$ on the boundary [see Eq.~\eqref{eq:ingoing defs}], we have
\begin{align}
    \langle \OO_1\!\!\left(\tfrac{t_1}{\lambda}\right) \OO_2\!\!\left(\tfrac{t_2}{\lambda}\right) \OO_3\!\!\left(\tfrac{t_3}{\lambda}\right) \OO_4\!\!\left(\tfrac{t_4}{\lambda}\right) \rangle 
    &= \Lambda \int \sqrt{-g} dV dR d^3 \vec{y}\, G_{\pd \rm B}^{1}\left(V- \tfrac{V_1}{\lambda},R\right) G_{\pd\rm B}^{2}\left( V- \tfrac{V_2}{\lambda},R\right) \nonumber \\
    & \qquad \qquad \qquad \qquad \qquad \times
   G_{\rm B \pd}^{3}\left( \tfrac{V_3}{\lambda} -V,R \right) G_{\rm B \pd}^{4}\left( \tfrac{ V_4}{\lambda}-V, R\right), \label{OOOO}
\end{align}
where we use the dimensionless radial coordinate
\begin{align}
    R := \frac{r-r_0}{r_0}.
\end{align}
We have now written out the dependence on $V$ and $R$ explicitly, but we still suppress the transverse spatial coordinates $\vec{y}$ and $\vec{y}_n$. 

Letting $\lambda \to 0$ enacts a flow to late times where the Aretakis growth appears.\footnote{Strictly speaking, we should also let $\vec{y}_n$ become large to enter in the regime where the full symmetry emerges.  However, these Euclidean spatial directions do not play an important role in the argument, so we leave this flow implicit, simply assuming that all space differences are in the regime \eqref{regime}.}  (Recall that we set $T=0$ for simplicity, considering the precisely extremal case where the growth persists indefinitely.) We therefore consider $\lambda$ to be small and seek the leading-in-$\lambda$ behavior of \eqref{OOOO}.  Dropping the middle factors for notational convenience, we have
\begin{align}
    \langle \OO_1\!\!\left(\tfrac{t_1}{\lambda}\right) \dots \OO_4\!\!\left(\tfrac{t_4}{\lambda}\right) \rangle 
    & = \Lambda \int \sqrt{-g} dV dR d^3 \vec{y}\, G_{\pd \rm B}^{1}\left(V- \tfrac{V_1}{\lambda},R\right) \dots 
    G_{\rm B \pd}^{4}\left( \tfrac{ V_4}{\lambda}-V, R\right), \nonumber \\
    & =\Lambda \int \sqrt{-g} \lambda^{-1} d\bar{V} dR d^3 \vec{y}\, G_{\pd \rm B}^{1}\left(\tfrac{\bar{V}-\bar{V}_1}{\lambda}, R \right) \dots G_{\rm B \pd}^{4}\left( \tfrac{\bar{V}_4-\bar{V}}{\lambda}, R \right), \nonumber \\
    & = \Lambda \int  \lambda^{-1} d\bar{V} d^3 \vec{y}\Bigg( \int_0^{\lambda^p} \sqrt{-g} dR G_{\pd \rm B}^{1}\left(\tfrac{\bar{V}-\bar{V}_1}{\lambda},R \right) \dots G_{\rm B \pd}^{4}\left( \tfrac{\bar{V}_4-\bar{V}}{\lambda},R  \right)  \nonumber \\ & \qquad \qquad \qquad \qquad \qquad \qquad  + \int_{\lambda^p}^\infty \sqrt{-g} dR G_{\pd \rm B}^{1}\left(\tfrac{\bar{V}-\bar{V}_1}{\lambda},R \right)  \dots G_{\rm B \pd}^{4}\left( \tfrac{\bar{V}_4-\bar{V}}{\lambda},R \right) \Bigg), \nonumber \\
    & = \Lambda \int \lambda^{-1} d\bar{V} d^3 \vec{y}\Bigg( \int_0^{\lambda^{p-1}} \sqrt{-g} \lambda d\bar{R} G_{\pd \rm B}^{1}\left(\tfrac{\bar{V}-\bar{V}_1}{\lambda},\lambda \bar{R} \right)  \dots G_{\rm B \pd}^{4}\left( \tfrac{\bar{V}_4-\bar{V}}{\lambda},\lambda \bar{R}  \right) \nonumber \\ & \qquad \qquad \qquad \qquad \qquad \qquad  + \int_{\lambda^p}^\infty \sqrt{-g} dR G_{\pd \rm B}^{1}\left(\tfrac{\bar{V}-\bar{V}_1}{\lambda}, R \right) \dots G_{\rm B \pd}^{4}\left( \tfrac{\bar{V}_4-\bar{V}}{\lambda},R  \right) \Bigg). \label{OOOO2}
\end{align}
We have introduced $\bar{R}=R/\lambda$ and $\bar{V}=\lambda V$.  We have also split the integral up at $R=\lambda^p$.  Although we make no explicit approximation yet, we choose  $0<p<1$ so that the split will be in the region of overlap $\lambda \ll R \ll 1$.

As $\lambda \to 0$ we enter the large-$\delta V$ regime of the Green function.  We are also assuming that only large $\delta y$ is important, which means that we may use the approximation \eqref{money}.  This expression has the following scaling behavior:
\begin{subequations}\label{emergent}
\begin{align}
    R \gg 1/\delta V \textrm{ (far-zone)}: & \qquad G_{\rm \pd B} \to \lambda \ G_{\rm \pd B} \textrm{ under } \delta V \to \lambda^{-1} \delta V. \\
    R \ll 1 \textrm{ (near-zone)}: & \qquad G_{\rm \pd B} \to \sqrt{\lambda} G_{\rm \pd B} \textrm{ under } \delta V \to \lambda^{-1} \delta V, \ R \to \lambda R.\label{emergent-near}
\end{align}
\end{subequations}
Although we do not study $G_{\rm B \pd}$ in detail, it is straightforward to establish that the same properties hold, where the rescaling of $R$ refers to the bulk point.  Using these properties in Eq.~\eqref{OOOO2} reveals that 
\begin{align}
    \langle \OO_1\!\!\left(\tfrac{t_1}{\lambda}\right) \dots \OO_4\!\!\left(\tfrac{t_4}{\lambda}\right) \rangle 
    & \approx \Lambda \int \lambda^{-1} d\bar{V} d^3 \vec{y}\Bigg( \int_0^{\lambda^{p-1}} \sqrt{-g} \lambda d\bar{R} \Big[ \sqrt{\lambda} G_{\pd \rm B}^{ 1 \rm near }\left(\bar{V}-\bar{V}_1, \bar{R} \right) \dots \sqrt{\lambda} G_{\rm B \pd}^{\rm 4 \rm near}\left( \bar{V}_4-\bar{V}, \bar{R}  \right)\Big]  \nonumber \\ & \qquad \qquad \qquad \qquad \qquad+ \int_{\lambda^p}^\infty \sqrt{-g}  dR \Big[ \lambda G_{\pd \rm B}^{1 \rm far}\left(\bar{V}-\bar{V}_1, R \right)  \dots \lambda G_{\rm B \pd}^{4 \rm far}\left( \bar{V}_4-\bar{V},R  \right)\Big] \Bigg), \label{OOOO <>< <>< fishies} \\
    & \approx \Lambda \int d\bar{V} d^3 \vec{y} \Bigg( \lambda^2 \int_0^{\infty} \sqrt{-g} d \bar{R} G_{\pd \rm B}^{ 1 \rm near} \dots  G_{\rm B \pd}^{4 \rm near} + \lambda^3 \int_0^{\infty} \sqrt{-g} d R G_{\pd \rm B}^{ 1 \rm far} \dots  G_{\rm B \pd}^{4 \rm far}  \Bigg), \label{OOOO red fish} \\
    & = \lambda^2 \frac{\Lambda r_0^3}{\ell^3} \int d\bar{V} d\bar{R} d^3 \vec{y} \ \! G_{\pd \rm B}^{ 1 \rm near} \dots  G_{\rm B \pd}^{4 \rm near} + O(\lambda^3). 
    \label{OOOO O_o}
\end{align}
In Eq.~\eqref{OOOO <>< <>< fishies} we have replaced the Green functions with their large-separation approximations (with near and far referring to the bottom and top lines of \eqref{money}, respectively).  In Eq.~\eqref{OOOO red fish} we note that the range of integration can be made infinite since the integrals separately converge.\footnote{The overlap region behavior of of $G_{\rm \pd B}$ is $1/\sqrt{R}$ (see Eq.~\eqref{money}).  We are assuming that the same behavior also occurs for $G_{B \pd}$.}).  In the last line we replace $\sqrt{-g}=(r/\ell)^3$ with its leading-in-$\lambda$ behavior and note that the far integral is subleading.  Equation~\eqref{OOOO O_o} shows that the leading, $\lambda^2$ piece of the four-point function is fixed entirely from near-zone quantities and hence scales like \eqref{4ptsym} under further rescalings of time.  This gives \eqref{4ptsym} the status of an emergent conformal symmetry \eqref{tconf} in the regime \eqref{regime}.

\begin{figure}
    \centering
    \includegraphics{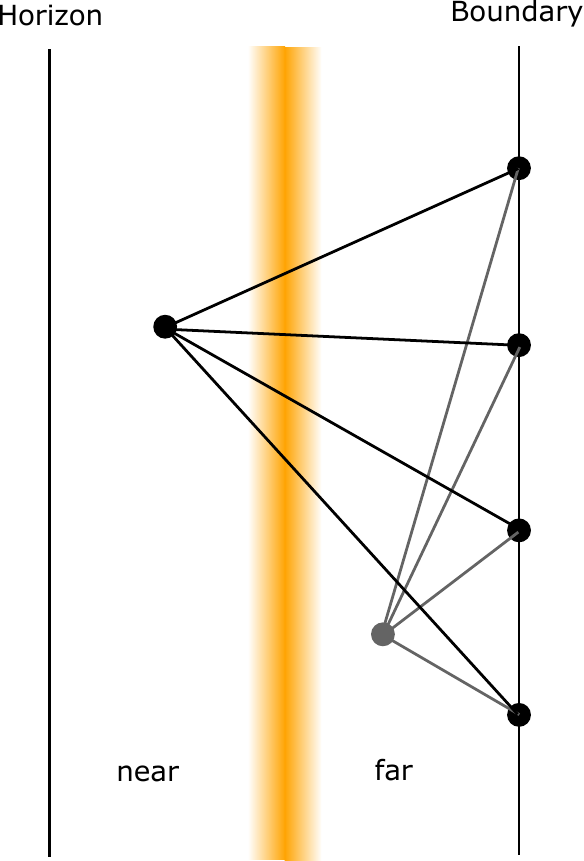} \qquad \qquad \qquad \qquad
    \includegraphics{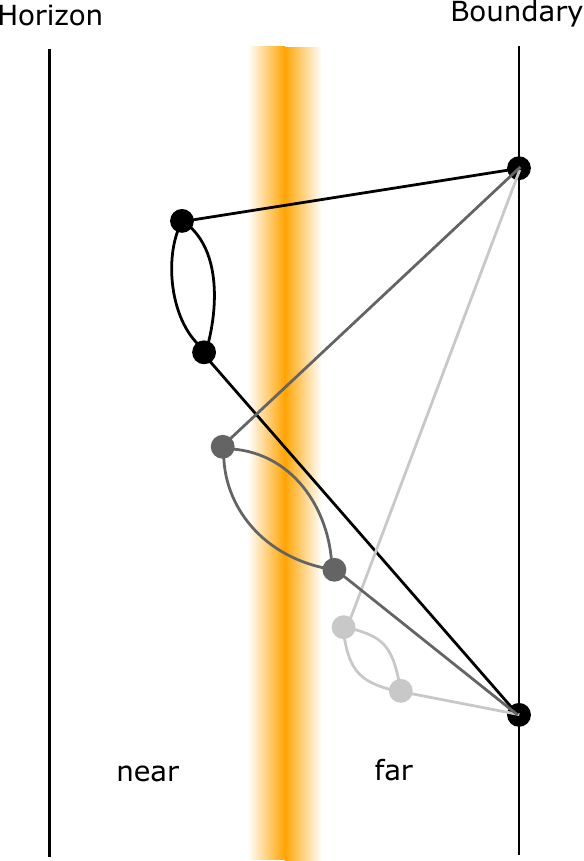}
    \caption{Diagrammatic representation of near-far splits in radial integrals for holographic correlators.  On the left, we show the leading dual four-point function of $\Phi^4$ theory, which is a tree-level diagram in the bulk.  On the right, we show the first $1/N$ correction to the dual propagator of $\Phi^3$ theory, which is a loop correction in the bulk.  The dominant contribution always comes from the graph with all vertices in the near-horizon region, shown here in the darkest color.}
    \label{fig:pics}
\end{figure}

\subsection{General argument and discussion}

Above we argued that the leading (tree-level) four-point function in $\Phi^4$ theory possesses an emergent conformal symmetry in the large-separation regime, ensuring its decay in time.  The key element is the split of the radial integral into near and far regimes, where powers of $\lambda$ can be counted.  The relevant counting is:
\begin{itemize}
\itemsep0em 
    \item $\lambda^0$ for each near-region vertex
    \item $\lambda^{-1}$ for each far-region vertex
    \item $\lambda^{1/2}$ for each boundary-near propagator ($G^{\rm near}_{\rm B \pd}$ or $G^{\rm near}_{\rm \pd B}$)
    \item $\lambda$ for each boundary-far propagator ($G^{\rm far}_{\rm B \pd}$ or $G^{\rm far}_{\rm \pd B}$)
\end{itemize}
The two integrals of Eq.~\eqref{OOOO <>< <>< fishies} are represented pictorially in the left panel of Fig.~\ref{fig:pics}.  The near integral is shown in black, while the far integral is shown in gray.  The counting above works as follows: The near integral receives $\lambda^0$ for the integration point and $(\lambda^{1/2})^4=\lambda^2$ for the propagators, for a total of $\lambda^2$.  The far integral receives $\lambda^{-1}$ for the far-region point and $\lambda^4$ for the propagators, for a total of $\lambda^3$.  The near-region therefore dominates, giving the expected $\lambda^2$.

For more general processes, the bulk-bulk propagator $G_{\rm BB}$ will appear as well.  In the large-separation regime this will have four different approximations ($G_{\rm ff}$, $G_{\rm fn}$, $G_{\rm nf}$, and $G_{\rm nn}$), depending on whether each radial point is in the near (n) or far (f) region.  Frequency-domain analysis makes clear that each approximation will have a scaling symmetry, where time is always rescaled ($\delta V \to \lambda^{-1} \delta V$) and radius is also rescaled $(R \to \lambda R)$ for the near point(s) only.  The scaling is a power of $\lambda^{1/2}$ for each far point: $G_{\rm nn}$ will be invariant, $G_{\rm fn}$ and $G_{\rm nf}$ will scale by $\lambda^{1/2}$, and $G_{\rm ff}$ will scale as $\lambda$.  The latter two scalings are reflected in Eqs.~\eqref{emergent} in the limit where the far point is taken to the boundary. 

In a given  diagram, each radial integral may be split into near and far-regions, which will again receive contributions as described above.  Combining this counting with the above scalings for the propagators gives the general counting as:
\begin{itemize}
\itemsep0em 
    \item $\lambda^{-1}$ for each dot in the far-region bulk (each far-region vertex)
    \item $\lambda^{1/2}$ for each line crossing the overlap region (near-far or far-near propagator)
    \item $\lambda$ for each line in the far-region (far-far propagator)
\end{itemize}
Dots and lines in the near-region (i.e., near-region vertices and near-near propagators) are given no $\lambda$ (i.e., $\lambda^0$).

For example, the right panel of Fig.~\ref{fig:pics} shows a $1/N$ correction to the dual two-point function in a $\Phi^3$ theory.  There are two integration points, whose three combinations are represented by different shades of gray.  The far-far combination (lightest color) receives $\lambda^{-1}$ for each of its two far-region vertices and $\lambda$ for each of its four far-far propagators, for a total of $\lambda^2$.  The far-near/near-far combination (intermediate color) receives $\lambda^{-1}$ for its single far-region vertex, $\lambda$ for its single far-far propagator, and $\lambda^{1/2}$ for each of its three near-far propagators, for a total of $\lambda^{3/2}$.  Finally, the near-near combination (darkest color) receives $\lambda^{1/2}$ for each of its two near-far propagators, for a total of $\lambda$.  This integral dominates, and the loop correction scales like $\lambda$.  This is the same scaling as the tree-level propagator \eqref{2ptsym} and respects the conformal symmetry \eqref{tconf}.

For a general diagram, it is clear that the contribution with all vertices in the near-region will always dominate in the same way.  For an $n$-point correlator $\langle \OO_1 \dots \OO_n \rangle$ at any order in the $1/N$ expansion, this dominant contribution contains $n$ propagators connecting the boundary to the near-region (each assigned $\lambda^{1/2}$) as well as an arbitrary number of near-horizon vertices and near-near propagators (all of which receive $\lambda^0$).  Thus the $n$-point correlator scales as $\lambda^{n/2}$ in the large-separation regime, precisely as required by the symmetry \eqref{tconf}.  These arguments hold for local non-derivative interactions in a scalar theory.  However, the arguments rely in essence only on symmetry, so we anticipate that similar arguments could be made in theories with additional fields and other local interactions.

It is interesting to note the key role played by the Aretakis instability in enforcing the conformal symmetry.  To illustrate, suppose that the near-region were somehow missing from the spacetime---that is, suppose that the large-separation approximation was uniformly given by the $1/t$ behavior of the far-region.  In this case only the lightest-colored integrals would exist in the diagrams of Fig.~\ref{fig:pics}.  This would make the leading $\Phi^4$ four-point function (left panel) scale as $\lambda^3$ instead of the $\lambda^2$ expected by symmetry.  Similarly, the $1/N$ correction to the $\Phi^3$ two-point function (right panel) would scale as $\lambda^2$ instead of the expected $\lambda$.  In general, the conformal symmetry would be broken.  A naive expectation might be that instabilities tend to break symmetries, but here we have one that enforces them!

We now return to the motivating question: What is the CFT dual to the Aretakis instability?  We have argued that dual correlation functions all decay, so there is no obvious boundary signature.  However, our explorations have revealed that the deep-bulk instability is intimately related to the emergent temporal conformal symmetry of the dual theory.  Could one consider the symmetry itself as somehow dual to the instability?  It is not clear exactly what this would mean, but it would certainly be interesting, as it would map a \textit{large} effect in the gravitational theory---unbounded stress-energy measured by infalling observers---to a bounded effect in the field theory.  Of course, another possibility is that there is no dual description at all: perhaps this phenomenon is just too deep in the bulk to be captured by holography.\footnote{However, note the Aretakis growth occurs outside the horizon, at distances $R \sim \sigma$, when the black hole is only nearly extremal.  One cannot therefore retreat to the position that holography applies only outside the horizon.}  This idea could be tested more sharply by studying the instability (or lack thereof) in a setting where the duality is believed to be complete.

We are forced to conclude that the holographic meaning of the Aretakis instability remains a mystery. 

\section*{Acknowledgements} We wish to thank Sean Hartnoll, Nabil Iqbal, and Don Marolf for helpful conversations.  This work was supported in part by NSF grant PHY-1506027 to the University of Arizona.  Portions of this work were completed at the Aspen Center for Physics, which is supported by NSF grant PHY-1607611.

\bibliography{MyReferences.bib}

\begin{thebibliography}{43}%
\makeatletter
\providecommand \@ifxundefined [1]{%
 \@ifx{#1\undefined}
}%
\providecommand \@ifnum [1]{%
 \ifnum #1\expandafter \@firstoftwo
 \else \expandafter \@secondoftwo
 \fi
}%
\providecommand \@ifx [1]{%
 \ifx #1\expandafter \@firstoftwo
 \else \expandafter \@secondoftwo
 \fi
}%
\providecommand \natexlab [1]{#1}%
\providecommand \enquote  [1]{``#1''}%
\providecommand \bibnamefont  [1]{#1}%
\providecommand \bibfnamefont [1]{#1}%
\providecommand \citenamefont [1]{#1}%
\providecommand \href@noop [0]{\@secondoftwo}%
\providecommand \href [0]{\begingroup \@sanitize@url \@href}%
\providecommand \@href[1]{\@@startlink{#1}\@@href}%
\providecommand \@@href[1]{\endgroup#1\@@endlink}%
\providecommand \@sanitize@url [0]{\catcode `\\12\catcode `\$12\catcode
  `\&12\catcode `\#12\catcode `\^12\catcode `\_12\catcode `\%12\relax}%
\providecommand \@@startlink[1]{}%
\providecommand \@@endlink[0]{}%
\providecommand \url  [0]{\begingroup\@sanitize@url \@url }%
\providecommand \@url [1]{\endgroup\@href {#1}{\urlprefix }}%
\providecommand \urlprefix  [0]{URL }%
\providecommand \Eprint [0]{\href }%
\providecommand \doibase [0]{http://dx.doi.org/}%
\providecommand \selectlanguage [0]{\@gobble}%
\providecommand \bibinfo  [0]{\@secondoftwo}%
\providecommand \bibfield  [0]{\@secondoftwo}%
\providecommand \translation [1]{[#1]}%
\providecommand \BibitemOpen [0]{}%
\providecommand \bibitemStop [0]{}%
\providecommand \bibitemNoStop [0]{.\EOS\space}%
\providecommand \EOS [0]{\spacefactor3000\relax}%
\providecommand \BibitemShut  [1]{\csname bibitem#1\endcsname}%
\let\auto@bib@innerbib\@empty
%</preamble>
\bibitem [{\citenamefont {Hartnoll}\ \emph {et~al.}(2016)\citenamefont
  {Hartnoll}, \citenamefont {Lucas},\ and\ \citenamefont
  {Sachdev}}]{Hartnoll:2016apf}%
  \BibitemOpen
  \bibfield  {author} {\bibinfo {author} {\bibfnamefont {S.~A.}\ \bibnamefont
  {Hartnoll}}, \bibinfo {author} {\bibfnamefont {A.}~\bibnamefont {Lucas}}, \
  and\ \bibinfo {author} {\bibfnamefont {S.}~\bibnamefont {Sachdev}},\
  }\href@noop {} {\  (\bibinfo {year} {2016})},\ \Eprint
  {http://arxiv.org/abs/1612.07324} {arXiv:1612.07324 [hep-th]} \BibitemShut
  {NoStop}%
%%CITATION = ARXIV:1612.07324;%%
\bibitem [{\citenamefont {Stewart}(2001)}]{strangemetals}%
  \BibitemOpen
  \bibfield  {author} {\bibinfo {author} {\bibfnamefont {G.~R.}\ \bibnamefont
  {Stewart}},\ }\href {\doibase 10.1103/RevModPhys.73.797} {\bibfield
  {journal} {\bibinfo  {journal} {Rev. Mod. Phys.}\ }\textbf {\bibinfo {volume}
  {73}},\ \bibinfo {pages} {797} (\bibinfo {year} {2001})}\BibitemShut
  {NoStop}%
\bibitem [{\citenamefont {Faulkner}\ \emph {et~al.}(2010)\citenamefont
  {Faulkner}, \citenamefont {Iqbal}, \citenamefont {Liu}, \citenamefont
  {McGreevy},\ and\ \citenamefont {Vegh}}]{Faulkner:2010zz}%
  \BibitemOpen
  \bibfield  {author} {\bibinfo {author} {\bibfnamefont {T.}~\bibnamefont
  {Faulkner}}, \bibinfo {author} {\bibfnamefont {N.}~\bibnamefont {Iqbal}},
  \bibinfo {author} {\bibfnamefont {H.}~\bibnamefont {Liu}}, \bibinfo {author}
  {\bibfnamefont {J.}~\bibnamefont {McGreevy}}, \ and\ \bibinfo {author}
  {\bibfnamefont {D.}~\bibnamefont {Vegh}},\ }\href {\doibase
  10.1126/science.1189134} {\bibfield  {journal} {\bibinfo  {journal}
  {Science}\ }\textbf {\bibinfo {volume} {329}},\ \bibinfo {pages} {1043}
  (\bibinfo {year} {2010})}\BibitemShut {NoStop}%
%%CITATION = SCIEA,329,1043;%%
\bibitem [{\citenamefont {Iqbal}\ \emph {et~al.}(2011)\citenamefont {Iqbal},
  \citenamefont {Liu},\ and\ \citenamefont {Mezei}}]{Iqbal:2011ae}%
  \BibitemOpen
  \bibfield  {author} {\bibinfo {author} {\bibfnamefont {N.}~\bibnamefont
  {Iqbal}}, \bibinfo {author} {\bibfnamefont {H.}~\bibnamefont {Liu}}, \ and\
  \bibinfo {author} {\bibfnamefont {M.}~\bibnamefont {Mezei}},\ }in\ \href
  {\doibase 10.1142/9789814350525_0013} {\emph {\bibinfo {booktitle}
  {{Proceedings, Theoretical Advanced Study Institute in Elementary Particle
  Physics (TASI 2010). String Theory and Its Applications: From meV to the
  Planck Scale: Boulder, Colorado, USA, June 1-25, 2010}}}}\ (\bibinfo {year}
  {2011})\ pp.\ \bibinfo {pages} {707--816},\ \Eprint
  {http://arxiv.org/abs/1110.3814} {arXiv:1110.3814 [hep-th]} \BibitemShut
  {NoStop}%
%%CITATION = ARXIV:1110.3814;%%
\bibitem [{\citenamefont {Aretakis}(2010)}]{Aretakis:2010gd}%
  \BibitemOpen
  \bibfield  {author} {\bibinfo {author} {\bibfnamefont {S.}~\bibnamefont
  {Aretakis}},\ }\href@noop {} {\  (\bibinfo {year} {2010})},\ \Eprint
  {http://arxiv.org/abs/1006.0283} {arXiv:1006.0283 [math.AP]} \BibitemShut
  {NoStop}%
%%CITATION = ARXIV:1006.0283;%%
\bibitem [{\citenamefont {Aretakis}(2011{\natexlab{a}})}]{Aretakis:2011ha}%
  \BibitemOpen
  \bibfield  {author} {\bibinfo {author} {\bibfnamefont {S.}~\bibnamefont
  {Aretakis}},\ }\href {\doibase 10.1007/s00220-011-1254-5} {\bibfield
  {journal} {\bibinfo  {journal} {Commun. Math. Phys.}\ }\textbf {\bibinfo
  {volume} {307}},\ \bibinfo {pages} {17} (\bibinfo {year}
  {2011}{\natexlab{a}})},\ \Eprint {http://arxiv.org/abs/1110.2007}
  {arXiv:1110.2007 [gr-qc]} \BibitemShut {NoStop}%
%%CITATION = ARXIV:1110.2007;%%
\bibitem [{\citenamefont {Aretakis}(2011{\natexlab{b}})}]{Aretakis:2011hc}%
  \BibitemOpen
  \bibfield  {author} {\bibinfo {author} {\bibfnamefont {S.}~\bibnamefont
  {Aretakis}},\ }\href {\doibase 10.1007/s00023-011-0110-7} {\bibfield
  {journal} {\bibinfo  {journal} {Annales Henri Poincare}\ }\textbf {\bibinfo
  {volume} {12}},\ \bibinfo {pages} {1491} (\bibinfo {year}
  {2011}{\natexlab{b}})},\ \Eprint {http://arxiv.org/abs/1110.2009}
  {arXiv:1110.2009 [gr-qc]} \BibitemShut {NoStop}%
%%CITATION = ARXIV:1110.2009;%%
\bibitem [{\citenamefont {Lucietti}\ \emph {et~al.}(2013)\citenamefont
  {Lucietti}, \citenamefont {Murata}, \citenamefont {Reall},\ and\
  \citenamefont {Tanahashi}}]{Lucietti:2012xr}%
  \BibitemOpen
  \bibfield  {author} {\bibinfo {author} {\bibfnamefont {J.}~\bibnamefont
  {Lucietti}}, \bibinfo {author} {\bibfnamefont {K.}~\bibnamefont {Murata}},
  \bibinfo {author} {\bibfnamefont {H.~S.}\ \bibnamefont {Reall}}, \ and\
  \bibinfo {author} {\bibfnamefont {N.}~\bibnamefont {Tanahashi}},\ }\href
  {\doibase 10.1007/JHEP03(2013)035} {\bibfield  {journal} {\bibinfo  {journal}
  {JHEP}\ }\textbf {\bibinfo {volume} {03}},\ \bibinfo {pages} {035} (\bibinfo
  {year} {2013})},\ \Eprint {http://arxiv.org/abs/1212.2557} {arXiv:1212.2557
  [gr-qc]} \BibitemShut {NoStop}%
%%CITATION = ARXIV:1212.2557;%%
\bibitem [{\citenamefont {Aretakis}(2013)}]{Aretakis:2012bm}%
  \BibitemOpen
  \bibfield  {author} {\bibinfo {author} {\bibfnamefont {S.}~\bibnamefont
  {Aretakis}},\ }\href {\doibase 10.1088/0264-9381/30/9/095010} {\bibfield
  {journal} {\bibinfo  {journal} {Class. Quant. Grav.}\ }\textbf {\bibinfo
  {volume} {30}},\ \bibinfo {pages} {095010} (\bibinfo {year} {2013})},\
  \Eprint {http://arxiv.org/abs/1212.1103} {arXiv:1212.1103 [gr-qc]}
  \BibitemShut {NoStop}%
%%CITATION = ARXIV:1212.1103;%%
\bibitem [{\citenamefont {Casals}\ \emph {et~al.}(2016)\citenamefont {Casals},
  \citenamefont {Gralla},\ and\ \citenamefont {Zimmerman}}]{Casals:2016mel}%
  \BibitemOpen
  \bibfield  {author} {\bibinfo {author} {\bibfnamefont {M.}~\bibnamefont
  {Casals}}, \bibinfo {author} {\bibfnamefont {S.~E.}\ \bibnamefont {Gralla}},
  \ and\ \bibinfo {author} {\bibfnamefont {P.}~\bibnamefont {Zimmerman}},\
  }\href {\doibase 10.1103/PhysRevD.94.064003} {\bibfield  {journal} {\bibinfo
  {journal} {Phys. Rev.}\ }\textbf {\bibinfo {volume} {D94}},\ \bibinfo {pages}
  {064003} (\bibinfo {year} {2016})},\ \Eprint
  {http://arxiv.org/abs/1606.08505} {arXiv:1606.08505 [gr-qc]} \BibitemShut
  {NoStop}%
%%CITATION = ARXIV:1606.08505;%%
\bibitem [{\citenamefont {Angelopoulos}\ \emph {et~al.}(2018)\citenamefont
  {Angelopoulos}, \citenamefont {Aretakis},\ and\ \citenamefont
  {Gajic}}]{Angelopoulos:2018uwb}%
  \BibitemOpen
  \bibfield  {author} {\bibinfo {author} {\bibfnamefont {Y.}~\bibnamefont
  {Angelopoulos}}, \bibinfo {author} {\bibfnamefont {S.}~\bibnamefont
  {Aretakis}}, \ and\ \bibinfo {author} {\bibfnamefont {D.}~\bibnamefont
  {Gajic}},\ }\href@noop {} {\  (\bibinfo {year} {2018})},\ \Eprint
  {http://arxiv.org/abs/1807.03802} {arXiv:1807.03802 [gr-qc]} \BibitemShut
  {NoStop}%
%%CITATION = ARXIV:1807.03802;%%
\bibitem [{\citenamefont {Murata}\ \emph {et~al.}(2013)\citenamefont {Murata},
  \citenamefont {Reall},\ and\ \citenamefont {Tanahashi}}]{Murata:2013daa}%
  \BibitemOpen
  \bibfield  {author} {\bibinfo {author} {\bibfnamefont {K.}~\bibnamefont
  {Murata}}, \bibinfo {author} {\bibfnamefont {H.~S.}\ \bibnamefont {Reall}}, \
  and\ \bibinfo {author} {\bibfnamefont {N.}~\bibnamefont {Tanahashi}},\ }\href
  {\doibase 10.1088/0264-9381/30/23/235007} {\bibfield  {journal} {\bibinfo
  {journal} {Class. Quant. Grav.}\ }\textbf {\bibinfo {volume} {30}},\ \bibinfo
  {pages} {235007} (\bibinfo {year} {2013})},\ \Eprint
  {http://arxiv.org/abs/1307.6800} {arXiv:1307.6800 [gr-qc]} \BibitemShut
  {NoStop}%
%%CITATION = ARXIV:1307.6800;%%
\bibitem [{\citenamefont {Gralla}\ \emph
  {et~al.}(2016{\natexlab{a}})\citenamefont {Gralla}, \citenamefont
  {Zimmerman},\ and\ \citenamefont {Zimmerman}}]{Gralla:2016sxp}%
  \BibitemOpen
  \bibfield  {author} {\bibinfo {author} {\bibfnamefont {S.~E.}\ \bibnamefont
  {Gralla}}, \bibinfo {author} {\bibfnamefont {A.}~\bibnamefont {Zimmerman}}, \
  and\ \bibinfo {author} {\bibfnamefont {P.}~\bibnamefont {Zimmerman}},\ }\href
  {\doibase 10.1103/PhysRevD.94.084017} {\bibfield  {journal} {\bibinfo
  {journal} {Phys. Rev.}\ }\textbf {\bibinfo {volume} {D94}},\ \bibinfo {pages}
  {084017} (\bibinfo {year} {2016}{\natexlab{a}})},\ \Eprint
  {http://arxiv.org/abs/1608.04739} {arXiv:1608.04739 [gr-qc]} \BibitemShut
  {NoStop}%
%%CITATION = ARXIV:1608.04739;%%
\bibitem [{\citenamefont {Zimmerman}(2017)}]{Zimmerman:2016qtn}%
  \BibitemOpen
  \bibfield  {author} {\bibinfo {author} {\bibfnamefont {P.}~\bibnamefont
  {Zimmerman}},\ }\href {\doibase 10.1103/PhysRevD.95.124032} {\bibfield
  {journal} {\bibinfo  {journal} {Phys. Rev.}\ }\textbf {\bibinfo {volume}
  {D95}},\ \bibinfo {pages} {124032} (\bibinfo {year} {2017})},\ \Eprint
  {http://arxiv.org/abs/1612.03172} {arXiv:1612.03172 [gr-qc]} \BibitemShut
  {NoStop}%
%%CITATION = ARXIV:1612.03172;%%
\bibitem [{\citenamefont {Aretakis}(2012)}]{aretakis2012decay}%
  \BibitemOpen
  \bibfield  {author} {\bibinfo {author} {\bibfnamefont {S.}~\bibnamefont
  {Aretakis}},\ }\href {\doibase 10.1016/j.jfa.2012.08.015} {\bibfield
  {journal} {\bibinfo  {journal} {J. Funct. Anal.}\ }\textbf {\bibinfo {volume}
  {263}},\ \bibinfo {pages} {2770} (\bibinfo {year} {2012})},\ \Eprint
  {http://arxiv.org/abs/1110.2006} {arXiv:1110.2006 [gr-qc]} \BibitemShut
  {NoStop}%
%%CITATION = ARXIV:1110.2006;%%
\bibitem [{\citenamefont {Aretakis}(2015)}]{Aretakis:2012ei}%
  \BibitemOpen
  \bibfield  {author} {\bibinfo {author} {\bibfnamefont {S.}~\bibnamefont
  {Aretakis}},\ }\href {\doibase 10.4310/ATMP.2015.v19.n3.a1} {\bibfield
  {journal} {\bibinfo  {journal} {Adv. Theor. Math. Phys.}\ }\textbf {\bibinfo
  {volume} {19}},\ \bibinfo {pages} {507} (\bibinfo {year} {2015})},\ \Eprint
  {http://arxiv.org/abs/1206.6598} {arXiv:1206.6598 [gr-qc]} \BibitemShut
  {NoStop}%
%%CITATION = ARXIV:1206.6598;%%
\bibitem [{\citenamefont {Lucietti}\ and\ \citenamefont
  {Reall}(2012)}]{Lucietti:2012sf}%
  \BibitemOpen
  \bibfield  {author} {\bibinfo {author} {\bibfnamefont {J.}~\bibnamefont
  {Lucietti}}\ and\ \bibinfo {author} {\bibfnamefont {H.~S.}\ \bibnamefont
  {Reall}},\ }\href {\doibase 10.1103/PhysRevD.86.104030} {\bibfield  {journal}
  {\bibinfo  {journal} {Phys. Rev.}\ }\textbf {\bibinfo {volume} {D86}},\
  \bibinfo {pages} {104030} (\bibinfo {year} {2012})},\ \Eprint
  {http://arxiv.org/abs/1208.1437} {arXiv:1208.1437 [gr-qc]} \BibitemShut
  {NoStop}%
%%CITATION = ARXIV:1208.1437;%%
\bibitem [{\citenamefont {Gralla}\ and\ \citenamefont
  {Zimmerman}(2018{\natexlab{a}})}]{gralla2018scaling}%
  \BibitemOpen
  \bibfield  {author} {\bibinfo {author} {\bibfnamefont {S.~E.}\ \bibnamefont
  {Gralla}}\ and\ \bibinfo {author} {\bibfnamefont {P.}~\bibnamefont
  {Zimmerman}},\ }\href@noop {} {\bibfield  {journal} {\bibinfo  {journal}
  {Journal of High Energy Physics}\ }\textbf {\bibinfo {volume} {2018}},\
  \bibinfo {pages} {61} (\bibinfo {year} {2018}{\natexlab{a}})}\BibitemShut
  {NoStop}%
\bibitem [{\citenamefont {Varma}\ \emph {et~al.}(1989)\citenamefont {Varma},
  \citenamefont {Littlewood}, \citenamefont {Schmitt-Rink}, \citenamefont
  {Abrahams},\ and\ \citenamefont {Ruckenstein}}]{Varma:1989zz}%
  \BibitemOpen
  \bibfield  {author} {\bibinfo {author} {\bibfnamefont {C.~M.}\ \bibnamefont
  {Varma}}, \bibinfo {author} {\bibfnamefont {P.~B.}\ \bibnamefont
  {Littlewood}}, \bibinfo {author} {\bibfnamefont {S.}~\bibnamefont
  {Schmitt-Rink}}, \bibinfo {author} {\bibfnamefont {E.}~\bibnamefont
  {Abrahams}}, \ and\ \bibinfo {author} {\bibfnamefont {A.~E.}\ \bibnamefont
  {Ruckenstein}},\ }\href {\doibase 10.1103/PhysRevLett.63.1996} {\bibfield
  {journal} {\bibinfo  {journal} {Phys. Rev. Lett.}\ }\textbf {\bibinfo
  {volume} {63}},\ \bibinfo {pages} {1996} (\bibinfo {year}
  {1989})}\BibitemShut {NoStop}%
%%CITATION = PRLTA,63,1996;%%
\bibitem [{\citenamefont {{Schr{\"o}der}}\ \emph {et~al.}(2000)\citenamefont
  {{Schr{\"o}der}}, \citenamefont {{Aeppli}}, \citenamefont {{Coldea}},
  \citenamefont {{Adams}}, \citenamefont {{Stockert}}, \citenamefont
  {{L{\"o}hneysen}}, \citenamefont {{Bucher}}, \citenamefont {{Ramazashvili}},\
  and\ \citenamefont {{Coleman}}}]{2000Natur.407..351S}%
  \BibitemOpen
  \bibfield  {author} {\bibinfo {author} {\bibfnamefont {A.}~\bibnamefont
  {{Schr{\"o}der}}}, \bibinfo {author} {\bibfnamefont {G.}~\bibnamefont
  {{Aeppli}}}, \bibinfo {author} {\bibfnamefont {R.}~\bibnamefont {{Coldea}}},
  \bibinfo {author} {\bibfnamefont {M.}~\bibnamefont {{Adams}}}, \bibinfo
  {author} {\bibfnamefont {O.}~\bibnamefont {{Stockert}}}, \bibinfo {author}
  {\bibfnamefont {H.~v.}\ \bibnamefont {{L{\"o}hneysen}}}, \bibinfo {author}
  {\bibfnamefont {E.}~\bibnamefont {{Bucher}}}, \bibinfo {author}
  {\bibfnamefont {R.}~\bibnamefont {{Ramazashvili}}}, \ and\ \bibinfo {author}
  {\bibfnamefont {P.}~\bibnamefont {{Coleman}}},\ }\href {\doibase
  10.1038/35030039} {\bibfield  {journal} {\bibinfo  {journal} {\nat}\ }\textbf
  {\bibinfo {volume} {407}},\ \bibinfo {pages} {351} (\bibinfo {year}
  {2000})},\ \Eprint {http://arxiv.org/abs/cond-mat/0011002} {cond-mat/0011002}
  \BibitemShut {NoStop}%
\bibitem [{\citenamefont {{Si}}\ \emph {et~al.}(2001)\citenamefont {{Si}},
  \citenamefont {{Rabello}}, \citenamefont {{Ingersent}},\ and\ \citenamefont
  {{Smith}}}]{2001Natur.413..804S}%
  \BibitemOpen
  \bibfield  {author} {\bibinfo {author} {\bibfnamefont {Q.}~\bibnamefont
  {{Si}}}, \bibinfo {author} {\bibfnamefont {S.}~\bibnamefont {{Rabello}}},
  \bibinfo {author} {\bibfnamefont {K.}~\bibnamefont {{Ingersent}}}, \ and\
  \bibinfo {author} {\bibfnamefont {J.~L.}\ \bibnamefont {{Smith}}},\ }\href
  {\doibase 10.1038/35101507} {\bibfield  {journal} {\bibinfo  {journal}
  {\nat}\ }\textbf {\bibinfo {volume} {413}},\ \bibinfo {pages} {804} (\bibinfo
  {year} {2001})},\ \Eprint {http://arxiv.org/abs/cond-mat/0011477}
  {cond-mat/0011477} \BibitemShut {NoStop}%
\bibitem [{\citenamefont {Iqbal}\ \emph {et~al.}(2012)\citenamefont {Iqbal},
  \citenamefont {Liu},\ and\ \citenamefont {Mezei}}]{Iqbal:2011in}%
  \BibitemOpen
  \bibfield  {author} {\bibinfo {author} {\bibfnamefont {N.}~\bibnamefont
  {Iqbal}}, \bibinfo {author} {\bibfnamefont {H.}~\bibnamefont {Liu}}, \ and\
  \bibinfo {author} {\bibfnamefont {M.}~\bibnamefont {Mezei}},\ }\href
  {\doibase 10.1007/JHEP04(2012)086} {\bibfield  {journal} {\bibinfo  {journal}
  {JHEP}\ }\textbf {\bibinfo {volume} {04}},\ \bibinfo {pages} {086} (\bibinfo
  {year} {2012})},\ \Eprint {http://arxiv.org/abs/1105.4621} {arXiv:1105.4621
  [hep-th]} \BibitemShut {NoStop}%
%%CITATION = ARXIV:1105.4621;%%
\bibitem [{\citenamefont {Gralla}\ and\ \citenamefont
  {Zimmerman}(2018{\natexlab{b}})}]{Gralla:2017lto}%
  \BibitemOpen
  \bibfield  {author} {\bibinfo {author} {\bibfnamefont {S.~E.}\ \bibnamefont
  {Gralla}}\ and\ \bibinfo {author} {\bibfnamefont {P.}~\bibnamefont
  {Zimmerman}},\ }\href {\doibase 10.1088/1361-6382/aab140} {\bibfield
  {journal} {\bibinfo  {journal} {Class. Quant. Grav.}\ }\textbf {\bibinfo
  {volume} {35}},\ \bibinfo {pages} {095002} (\bibinfo {year}
  {2018}{\natexlab{b}})},\ \Eprint {http://arxiv.org/abs/1711.00855}
  {arXiv:1711.00855 [gr-qc]} \BibitemShut {NoStop}%
%%CITATION = ARXIV:1711.00855;%%
\bibitem [{\citenamefont {Hadar}\ and\ \citenamefont
  {Reall}(2017)}]{Hadar:2017ven}%
  \BibitemOpen
  \bibfield  {author} {\bibinfo {author} {\bibfnamefont {S.}~\bibnamefont
  {Hadar}}\ and\ \bibinfo {author} {\bibfnamefont {H.~S.}\ \bibnamefont
  {Reall}},\ }\href {\doibase 10.1007/JHEP12(2017)062} {\bibfield  {journal}
  {\bibinfo  {journal} {JHEP}\ }\textbf {\bibinfo {volume} {12}},\ \bibinfo
  {pages} {062} (\bibinfo {year} {2017})},\ \Eprint
  {http://arxiv.org/abs/1709.09668} {arXiv:1709.09668 [hep-th]} \BibitemShut
  {NoStop}%
%%CITATION = ARXIV:1709.09668;%%
\bibitem [{\citenamefont {Burko}\ and\ \citenamefont
  {Khanna}(2018)}]{Burko:2017eky}%
  \BibitemOpen
  \bibfield  {author} {\bibinfo {author} {\bibfnamefont {L.~M.}\ \bibnamefont
  {Burko}}\ and\ \bibinfo {author} {\bibfnamefont {G.}~\bibnamefont {Khanna}},\
  }\href {\doibase 10.1103/PhysRevD.97.061502} {\bibfield  {journal} {\bibinfo
  {journal} {Phys. Rev.}\ }\textbf {\bibinfo {volume} {D97}},\ \bibinfo {pages}
  {061502} (\bibinfo {year} {2018})},\ \Eprint
  {http://arxiv.org/abs/1709.10155} {arXiv:1709.10155 [gr-qc]} \BibitemShut
  {NoStop}%
%%CITATION = ARXIV:1709.10155;%%
\bibitem [{\citenamefont {Gralla}\ and\ \citenamefont
  {Zimmerman}(2018{\natexlab{c}})}]{Gralla:2018xzo}%
  \BibitemOpen
  \bibfield  {author} {\bibinfo {author} {\bibfnamefont {S.~E.}\ \bibnamefont
  {Gralla}}\ and\ \bibinfo {author} {\bibfnamefont {P.}~\bibnamefont
  {Zimmerman}},\ }\href {\doibase 10.1007/JHEP06(2018)061} {\bibfield
  {journal} {\bibinfo  {journal} {JHEP}\ }\textbf {\bibinfo {volume} {06}},\
  \bibinfo {pages} {061} (\bibinfo {year} {2018}{\natexlab{c}})},\ \Eprint
  {http://arxiv.org/abs/1804.04753} {arXiv:1804.04753 [gr-qc]} \BibitemShut
  {NoStop}%
%%CITATION = ARXIV:1804.04753;%%
\bibitem [{\citenamefont {Chamblin}\ \emph {et~al.}(1999)\citenamefont
  {Chamblin}, \citenamefont {Emparan}, \citenamefont {Johnson},\ and\
  \citenamefont {Myers}}]{Chamblin:1999tk}%
  \BibitemOpen
  \bibfield  {author} {\bibinfo {author} {\bibfnamefont {A.}~\bibnamefont
  {Chamblin}}, \bibinfo {author} {\bibfnamefont {R.}~\bibnamefont {Emparan}},
  \bibinfo {author} {\bibfnamefont {C.~V.}\ \bibnamefont {Johnson}}, \ and\
  \bibinfo {author} {\bibfnamefont {R.~C.}\ \bibnamefont {Myers}},\ }\href
  {\doibase 10.1103/PhysRevD.60.064018} {\bibfield  {journal} {\bibinfo
  {journal} {Phys. Rev.}\ }\textbf {\bibinfo {volume} {D60}},\ \bibinfo {pages}
  {064018} (\bibinfo {year} {1999})},\ \Eprint
  {http://arxiv.org/abs/hep-th/9902170} {arXiv:hep-th/9902170 [hep-th]}
  \BibitemShut {NoStop}%
%%CITATION = HEP-TH/9902170;%%
\bibitem [{\citenamefont {Faulkner}\ \emph
  {et~al.}(2011{\natexlab{a}})\citenamefont {Faulkner}, \citenamefont {Liu},
  \citenamefont {McGreevy},\ and\ \citenamefont {Vegh}}]{Faulkner:2009wj}%
  \BibitemOpen
  \bibfield  {author} {\bibinfo {author} {\bibfnamefont {T.}~\bibnamefont
  {Faulkner}}, \bibinfo {author} {\bibfnamefont {H.}~\bibnamefont {Liu}},
  \bibinfo {author} {\bibfnamefont {J.}~\bibnamefont {McGreevy}}, \ and\
  \bibinfo {author} {\bibfnamefont {D.}~\bibnamefont {Vegh}},\ }\href {\doibase
  10.1103/PhysRevD.83.125002} {\bibfield  {journal} {\bibinfo  {journal} {Phys.
  Rev.}\ }\textbf {\bibinfo {volume} {D83}},\ \bibinfo {pages} {125002}
  (\bibinfo {year} {2011}{\natexlab{a}})},\ \Eprint
  {http://arxiv.org/abs/0907.2694} {arXiv:0907.2694 [hep-th]} \BibitemShut
  {NoStop}%
%%CITATION = ARXIV:0907.2694;%%
\bibitem [{\citenamefont {Son}\ and\ \citenamefont
  {Starinets}(2002)}]{Son:2002sd}%
  \BibitemOpen
  \bibfield  {author} {\bibinfo {author} {\bibfnamefont {D.~T.}\ \bibnamefont
  {Son}}\ and\ \bibinfo {author} {\bibfnamefont {A.~O.}\ \bibnamefont
  {Starinets}},\ }\href {\doibase 10.1088/1126-6708/2002/09/042} {\bibfield
  {journal} {\bibinfo  {journal} {JHEP}\ }\textbf {\bibinfo {volume} {09}},\
  \bibinfo {pages} {042} (\bibinfo {year} {2002})},\ \Eprint
  {http://arxiv.org/abs/hep-th/0205051} {arXiv:hep-th/0205051 [hep-th]}
  \BibitemShut {NoStop}%
%%CITATION = HEP-TH/0205051;%%
\bibitem [{\citenamefont {Ren}(2012)}]{Ren:2012hg}%
  \BibitemOpen
  \bibfield  {author} {\bibinfo {author} {\bibfnamefont {J.}~\bibnamefont
  {Ren}},\ }\href@noop {} {\  (\bibinfo {year} {2012})},\ \Eprint
  {http://arxiv.org/abs/1210.2722} {arXiv:1210.2722 [hep-th]} \BibitemShut
  {NoStop}%
%%CITATION = ARXIV:1210.2722;%%
\bibitem [{\citenamefont {Ren}(2013)}]{ren2013quantum}%
  \BibitemOpen
  \bibfield  {author} {\bibinfo {author} {\bibfnamefont {J.}~\bibnamefont
  {Ren}},\ }\emph {\bibinfo {title} {Quantum Critical Systems from AdS/CFT}},\
  \href@noop {} {Ph.D. thesis},\ \bibinfo  {school} {Princeton University}
  (\bibinfo {year} {2013})\BibitemShut {NoStop}%
\bibitem [{{\relax DLMF}()}]{NIST:DLMF}%
  \BibitemOpen
  {\relax DLMF},\ \href {http://dlmf.nist.gov/} {\enquote {\bibinfo {title}
  {{NIST Digital Library of Mathematical Functions}},}\ }\bibinfo
  {howpublished} {http://dlmf.nist.gov/, Release 1.0.5 of 2012-10-01},\
  \bibinfo {note} {online companion to \cite{nist}}\BibitemShut {NoStop}%
\bibitem [{\citenamefont {Iqbal}\ \emph {et~al.}(2010)\citenamefont {Iqbal},
  \citenamefont {Liu}, \citenamefont {Mezei},\ and\ \citenamefont
  {Si}}]{Iqbal:2010eh}%
  \BibitemOpen
  \bibfield  {author} {\bibinfo {author} {\bibfnamefont {N.}~\bibnamefont
  {Iqbal}}, \bibinfo {author} {\bibfnamefont {H.}~\bibnamefont {Liu}}, \bibinfo
  {author} {\bibfnamefont {M.}~\bibnamefont {Mezei}}, \ and\ \bibinfo {author}
  {\bibfnamefont {Q.}~\bibnamefont {Si}},\ }\href {\doibase
  10.1103/PhysRevD.82.045002} {\bibfield  {journal} {\bibinfo  {journal} {Phys.
  Rev.}\ }\textbf {\bibinfo {volume} {D82}},\ \bibinfo {pages} {045002}
  (\bibinfo {year} {2010})},\ \Eprint {http://arxiv.org/abs/1003.0010}
  {arXiv:1003.0010 [hep-th]} \BibitemShut {NoStop}%
%%CITATION = ARXIV:1003.0010;%%
\bibitem [{\citenamefont {Denef}\ and\ \citenamefont
  {Hartnoll}(2009)}]{Denef:2009tp}%
  \BibitemOpen
  \bibfield  {author} {\bibinfo {author} {\bibfnamefont {F.}~\bibnamefont
  {Denef}}\ and\ \bibinfo {author} {\bibfnamefont {S.~A.}\ \bibnamefont
  {Hartnoll}},\ }\href {\doibase 10.1103/PhysRevD.79.126008} {\bibfield
  {journal} {\bibinfo  {journal} {Phys. Rev.}\ }\textbf {\bibinfo {volume}
  {D79}},\ \bibinfo {pages} {126008} (\bibinfo {year} {2009})},\ \Eprint
  {http://arxiv.org/abs/0901.1160} {arXiv:0901.1160 [hep-th]} \BibitemShut
  {NoStop}%
%%CITATION = ARXIV:0901.1160;%%
\bibitem [{\citenamefont {Gralla}\ \emph
  {et~al.}(2016{\natexlab{b}})\citenamefont {Gralla}, \citenamefont {Hughes},\
  and\ \citenamefont {Warburton}}]{Gralla:2016qfw}%
  \BibitemOpen
  \bibfield  {author} {\bibinfo {author} {\bibfnamefont {S.~E.}\ \bibnamefont
  {Gralla}}, \bibinfo {author} {\bibfnamefont {S.~A.}\ \bibnamefont {Hughes}},
  \ and\ \bibinfo {author} {\bibfnamefont {N.}~\bibnamefont {Warburton}},\
  }\href {\doibase 10.1088/0264-9381/33/15/155002} {\bibfield  {journal}
  {\bibinfo  {journal} {Class. Quant. Grav.}\ }\textbf {\bibinfo {volume}
  {33}},\ \bibinfo {pages} {155002} (\bibinfo {year} {2016}{\natexlab{b}})},\
  \Eprint {http://arxiv.org/abs/1603.01221} {arXiv:1603.01221 [gr-qc]}
  \BibitemShut {NoStop}%
%%CITATION = ARXIV:1603.01221;%%
\bibitem [{\citenamefont {Comp{\`e}re}\ \emph {et~al.}(2017)\citenamefont
  {Comp{\`e}re}, \citenamefont {Fransen}, \citenamefont {Hertog},\ and\
  \citenamefont {Long}}]{Compere:2017hsi}%
  \BibitemOpen
  \bibfield  {author} {\bibinfo {author} {\bibfnamefont {G.}~\bibnamefont
  {Comp{\`e}re}}, \bibinfo {author} {\bibfnamefont {K.}~\bibnamefont
  {Fransen}}, \bibinfo {author} {\bibfnamefont {T.}~\bibnamefont {Hertog}}, \
  and\ \bibinfo {author} {\bibfnamefont {J.}~\bibnamefont {Long}},\ }\href@noop
  {} {\  (\bibinfo {year} {2017})},\ \Eprint {http://arxiv.org/abs/1712.07130}
  {arXiv:1712.07130 [gr-qc]} \BibitemShut {NoStop}%
%%CITATION = ARXIV:1712.07130;%%
\bibitem [{\citenamefont {Project}\ \emph {et~al.}(1954)\citenamefont
  {Project}, \citenamefont {Bateman}, \citenamefont {Erd{\'e}lyi},\ and\
  \citenamefont {of~Naval~Research}}]{bateman1954tables}%
  \BibitemOpen
  \bibfield  {author} {\bibinfo {author} {\bibfnamefont {B.~M.}\ \bibnamefont
  {Project}}, \bibinfo {author} {\bibfnamefont {H.}~\bibnamefont {Bateman}},
  \bibinfo {author} {\bibfnamefont {A.}~\bibnamefont {Erd{\'e}lyi}}, \ and\
  \bibinfo {author} {\bibfnamefont {U.~S.~O.}\ \bibnamefont
  {of~Naval~Research}},\ }\href@noop {} {\emph {\bibinfo {title} {Tables of
  Integral Transforms: Based, in Part, on Notes Left by Harry Bateman}}},\
  \bibinfo {series} {Tables of Integral Transforms: Based, in Part, on Notes
  Left by Harry Bateman}\ No.\ \bibinfo {number} {v. 2}\ (\bibinfo  {publisher}
  {McGraw-Hill},\ \bibinfo {year} {1954})\BibitemShut {NoStop}%
\bibitem [{\citenamefont {Faulkner}\ \emph
  {et~al.}(2011{\natexlab{b}})\citenamefont {Faulkner}, \citenamefont {Iqbal},
  \citenamefont {Liu}, \citenamefont {McGreevy},\ and\ \citenamefont
  {Vegh}}]{Faulkner:2011tm}%
  \BibitemOpen
  \bibfield  {author} {\bibinfo {author} {\bibfnamefont {T.}~\bibnamefont
  {Faulkner}}, \bibinfo {author} {\bibfnamefont {N.}~\bibnamefont {Iqbal}},
  \bibinfo {author} {\bibfnamefont {H.}~\bibnamefont {Liu}}, \bibinfo {author}
  {\bibfnamefont {J.}~\bibnamefont {McGreevy}}, \ and\ \bibinfo {author}
  {\bibfnamefont {D.}~\bibnamefont {Vegh}},\ }\href {\doibase
  10.1098/rsta.2010.0354} {\bibfield  {journal} {\bibinfo  {journal} {Phil.
  Trans. Roy. Soc.}\ }\textbf {\bibinfo {volume} {A 369}},\ \bibinfo {pages}
  {1640} (\bibinfo {year} {2011}{\natexlab{b}})},\ \Eprint
  {http://arxiv.org/abs/1101.0597} {arXiv:1101.0597 [hep-th]} \BibitemShut
  {NoStop}%
%%CITATION = ARXIV:1101.0597;%%
\bibitem [{\citenamefont {Hod}(2008)}]{Hod:2008zz}%
  \BibitemOpen
  \bibfield  {author} {\bibinfo {author} {\bibfnamefont {S.}~\bibnamefont
  {Hod}},\ }\href {\doibase 10.1103/PhysRevD.78.084035} {\bibfield  {journal}
  {\bibinfo  {journal} {Phys. Rev.}\ }\textbf {\bibinfo {volume} {D78}},\
  \bibinfo {pages} {084035} (\bibinfo {year} {2008})},\ \Eprint
  {http://arxiv.org/abs/0811.3806} {arXiv:0811.3806 [gr-qc]} \BibitemShut
  {NoStop}%
%%CITATION = ARXIV:0811.3806;%%
\bibitem [{\citenamefont {Yang}\ \emph {et~al.}(2013)\citenamefont {Yang},
  \citenamefont {Zimmerman}, \citenamefont {Zengino{\u g}lu}, \citenamefont
  {Zhang}, \citenamefont {Berti} \emph {et~al.}}]{Yang:2013uba}%
  \BibitemOpen
  \bibfield  {author} {\bibinfo {author} {\bibfnamefont {H.}~\bibnamefont
  {Yang}}, \bibinfo {author} {\bibfnamefont {A.}~\bibnamefont {Zimmerman}},
  \bibinfo {author} {\bibfnamefont {A.}~\bibnamefont {Zengino{\u g}lu}},
  \bibinfo {author} {\bibfnamefont {F.}~\bibnamefont {Zhang}}, \bibinfo
  {author} {\bibfnamefont {E.}~\bibnamefont {Berti}},  \emph {et~al.},\ }\href
  {\doibase 10.1103/PhysRevD.88.044047} {\bibfield  {journal} {\bibinfo
  {journal} {Phys.Rev.}\ }\textbf {\bibinfo {volume} {D88}},\ \bibinfo {pages}
  {044047} (\bibinfo {year} {2013})},\ \Eprint {http://arxiv.org/abs/1307.8086}
  {arXiv:1307.8086 [gr-qc]} \BibitemShut {NoStop}%
%%CITATION = ARXIV:1307.8086;%%
\bibitem [{\citenamefont {Banks}\ \emph {et~al.}(1998)\citenamefont {Banks},
  \citenamefont {Douglas}, \citenamefont {Horowitz},\ and\ \citenamefont
  {Martinec}}]{Banks:1998dd}%
  \BibitemOpen
  \bibfield  {author} {\bibinfo {author} {\bibfnamefont {T.}~\bibnamefont
  {Banks}}, \bibinfo {author} {\bibfnamefont {M.~R.}\ \bibnamefont {Douglas}},
  \bibinfo {author} {\bibfnamefont {G.~T.}\ \bibnamefont {Horowitz}}, \ and\
  \bibinfo {author} {\bibfnamefont {E.~J.}\ \bibnamefont {Martinec}},\
  }\href@noop {} {\  (\bibinfo {year} {1998})},\ \Eprint
  {http://arxiv.org/abs/hep-th/9808016} {arXiv:hep-th/9808016 [hep-th]}
  \BibitemShut {NoStop}%
%%CITATION = HEP-TH/9808016;%%
\bibitem [{\citenamefont {Harlow}\ and\ \citenamefont
  {Stanford}(2011)}]{Harlow:2011ke}%
  \BibitemOpen
  \bibfield  {author} {\bibinfo {author} {\bibfnamefont {D.}~\bibnamefont
  {Harlow}}\ and\ \bibinfo {author} {\bibfnamefont {D.}~\bibnamefont
  {Stanford}},\ }\href@noop {} {\  (\bibinfo {year} {2011})},\ \Eprint
  {http://arxiv.org/abs/1104.2621} {arXiv:1104.2621 [hep-th]} \BibitemShut
  {NoStop}%
%%CITATION = ARXIV:1104.2621;%%
\bibitem [{\citenamefont {Olver}\ \emph {et~al.}(2010)\citenamefont {Olver},
  \citenamefont {Lozier}, \citenamefont {Boisvert},\ and\ \citenamefont
  {Clark}}]{nist}%
  \BibitemOpen
  \bibfield  {author} {\bibinfo {author} {\bibfnamefont {F.~W.}\ \bibnamefont
  {Olver}}, \bibinfo {author} {\bibfnamefont {D.~W.}\ \bibnamefont {Lozier}},
  \bibinfo {author} {\bibfnamefont {R.~F.}\ \bibnamefont {Boisvert}}, \ and\
  \bibinfo {author} {\bibfnamefont {C.~W.}\ \bibnamefont {Clark}},\ }\href
  {http://dlmf.nist.gov/} {\emph {\bibinfo {title} {NIST Handbook of
  Mathematical Functions}}},\ \bibinfo {edition} {1st}\ ed.\ (\bibinfo
  {publisher} {Cambridge University Press},\ \bibinfo {address} {New York, NY,
  USA},\ \bibinfo {year} {2010})\BibitemShut {NoStop}%
\end{thebibliography}%
\end{document}